\documentclass[review]{elsarticle}
\usepackage{amsthm}
\usepackage{bbm}
\usepackage{pifont}
\usepackage{bbding}
\usepackage{xcolor}
\usepackage{mathrsfs}
\usepackage{epsfig,amsmath,amssymb,graphicx}
\usepackage{epstopdf}
\usepackage{lineno}
\usepackage{geometry}
\usepackage[colorlinks, linkcolor=blue, anchorcolor=blue,
citecolor=blue]{hyperref}
\usepackage{caption}
\usepackage{graphicx}
\usepackage{subfigure}
\newtheorem{remark}{Remark}

\usepackage{lineno,hyperref}

\journal{Nonlinear Analysis: Modelling and Control}

\begin{document}

\begin{frontmatter}

\title{Pattern formation in a predator-prey model with Allee effect and hyperbolic mortality on networked and non-networked environments}


\author[mymainaddress,mysecondaryaddress]{Yong Ye}
\ead{13339239813@163.com}

\author[mymainaddress,mysecondaryaddress]{Jiaying Zhou\corref{mycorrespondingauthor}}
\cortext[mycorrespondingauthor]{Corresponding author}
\ead{jyzhou0513@gmail.com}

\address[mymainaddress]{School of Science, Harbin Institute of Technology, Shenzhen, 518055/P. R. China}
\address[mysecondaryaddress]{Departament d'Enginyeria Inform\`{a}tica i Matem\`{a}tiques, Universitat Rovira i Virgili, E-43007 Tarragona, Spain}

\begin{abstract}
With the development of network science, Turing pattern has been proven to be formed in discrete media such as complex networks, opening up the possibility of exploring it as a generation mechanism in the context of biology, chemistry, and physics. Turing instability in the predator-prey system has been widely studied in recent years. We hope to use the predator-prey interaction relationship in biological populations to explain the influence of network topology on pattern formation. In this paper, we establish a predator-prey model with weak Allee effect, analyze and verify the Turing instability conditions on the large ER (Erd\"{o}s-R\'{e}nyi) random network with the help of Turing stability theory and numerical experiments, and obtain the Turing instability region. \textcolor{red}{The results indicate that diffusion plays a decisive role in the generation of spatial patterns, whether in continuous or discrete media. For spatiotemporal patterns, different initial values can also bring about changes in the pattern.} When we analyze the model based on the network framework, we find that the average degree of the network has an important impact on the model, and different average degrees will lead to changes in the distribution pattern of the population.
\end{abstract}
\begin{keyword}
Turing pattern\sep predator-prey\sep random network\sep average degree
\MSC[2010] 92D25\sep 35K57
\end{keyword}
\end{frontmatter}

\section{Introduction}\label{section1}
Since the pioneering work of Lotka and Volterra, the dynamic characteristics between predator and prey populations have always been an important research topic in mathematical biology~\cite{murray2002mathematical}. After that, many researchers worked to improve the model on this basis~\cite{nakao2010turing,nagano2012phase,zhang2014spatio,liu2019pattern,ye2019dynamic,ye2021bifurcation,ye2022promotion}. \textcolor{red}{In 2012, Nagano and Maeda explored the spatial distribution of predators and prey using the well-known model of Rosenzweig and MacArthur. They gave the phase diagram of this predator-prey model and studied the lattice formation on this model~\cite{nagano2012phase}. However, in 2014, Zhang and his collaborators found that there would be no Turing pattern if only the death term of the predator was represented by a linear function. Therefore, they studied the pattern formation on the predator-prey model with hyperbolic mortality by selecting appropriate control parameters and found many interesting patterns, such as hexagonal spots and stripe patterns~\cite{zhang2014spatio}. Allee effect widely exists in biological systems and is often considered by researchers in their established models. Inspired by this, Liu et al. introduced weak Allee effect into the prey population based on previous studies. They focused on studying the Allee effect on the spatial distribution of species and found that Allee effect increases the isolation of spatial patches. From a biological perspective, Allee effect make the spatial distribution of the population more concentrated, which is beneficial for the continuation of the species~\cite{liu2019pattern}. This model is the scenario mentioned in this paper for non-networked environments as follows
\begin{equation}\label{liu_model}
\left\{\begin{array}{l}
\frac{\partial u}{\partial t}-d_1 \Delta u=\alpha u(1-u)\left(\frac{u}{u+A}\right)-\frac{\alpha u v}{1+\beta u}, \\
\frac{\partial v}{\partial t}-d_2 \Delta v=\frac{\beta u v}{1+\beta u}-\frac{\gamma v^2}{e+\eta v}, \\
\frac{\partial u}{\partial n}=\frac{\partial v}{\partial n}=0,(x, y) \in \partial \Omega, t>0, \\
u(0, x, y)=u_0(x, y) \geq 0, v(0, x, y)=v_0(x, y) \geq 0 .
\end{array}\right.
\end{equation}}
As we know, reaction-diffusion systems support much complex self-organization phenomenons~\cite{turing1952chemical}. The research on reaction-diffusion systems in continuous media, especially about Turing pattern has been well-developed. However, in discrete media, such as complex networks, there are many unknown possibilities. With the rapid development of network science, researchers began to try to explore more possibilities of such reaction-diffusion systems on the network platform. As early as 1971, Othmer and Scriven tried to study the influence of network structure on Turing instability on several regular planar and polyhedral networks~\cite{othmer1971instability}. Subsequently, a series of related works appeared, but most of them were based on regular lattice networks or small networks~\cite{othmer1974non,horsthemke2004network,moore2005localized}. In 2010, Nakao and Mikhailov pointed out this problem. They took the Mimura-Murray model in the predator-prey population as an example to study the pattern formation on large random networks. The results showed that Turing instability would lead to spontaneous differentiation of network nodes into rich and weak activator groups, and multiple steady-state coexistences and hysteresis effects were observed~\cite{nakao2010turing}. After that, pattern formation on complex networks considering different backgrounds has been studied by researchers~\cite{fernandes2012turing,asllani2014theory,asllani2014turing,kouvaris2015pattern,petit2016pattern,petit2017theory,hu2021turing,zhou2022bifurcation,zhou2022complex,zhu2022pattern,ma2023complex,xie2021transmission,wang2022bifurcation,ye2023pattern,zhou2023pattern}. In 2019, Chang and his collaborators tried to introduce delay into the Leslie-Gower model and studied the effect of delay on the shape of the pattern based on several regular networks. The results showed that introducing delay would bring about many beautiful patterns~\cite{chang2019delay}. In the following year, he and his colleagues also studied the Turing pattern on the multiplex network, taking into account the case of self-diffusion and cross-diffusion, and also found patterns with rich characteristics~\cite{gao2020cross}. Over the years, they explored the pattern generation in different regular networks, random networks, and multiplex networks based on the predator-prey model and SIR model respectively~\cite{chang2020cross,liu2020turing,chang2022optimal,liu2022optimal,gao2022optimal}. Similarly, Zheng and his colleagues have done a lot of interesting work on pattern generation on the network in recent years~\cite{zheng2020turing,zheng2022turing,zheng2022spatiotemporal,chen2022hopf}. In addition, Tian and his collaborators have also carried out research on the mathematical theory of reaction-diffusion systems based on complex networks~\cite{tian2019pattern,liu2020weighted,liu2020network,gan2020delay,tian2023asymptotic}. \textcolor{red}{The reaction-diffusion process under consideration is characterized by discrete media as opposed to continuous media, as Liu et al. pointed out in 2020. Typically, species are dispersed in various areas, which can be visualized as a complex network~\cite{liu2020turing}. Inspired by this idea, we hope to consider the discrete media instead of the reaction-diffusion model on the continuous media to study Turing pattern on the predator-prey model with weak Allee effect under the complex network and try to explore the influence of network topology on the pattern formation. Therefore, a predator-prey model with Allee effect and hyperbolic mortality in networked environments can be described as}
\begin{equation}\label{self-diffusion}
\begin{cases}\frac{\mathbf{d} u_i}{\mathbf{d} t}=f\left(u_i, v_i\right)+d_{1} \sum_{j=1}^N L_{i j} u_j,  \\
\frac{\mathbf{d} v_i}{\mathbf{d} t}=g\left(u_i, v_i\right)+d_{2} \sum_{j=1}^N L_{i j} v_j\\
u_i(0)\geq 0,~v_i(0)\geq 0.\end{cases}
\end{equation}
The reaction term is expressed in the following form:
$$
f(u_i, v_i)=\alpha u_i(1-u_i)\left(\frac{u_i}{u_i+A}\right)-\frac{\alpha u_i v_i}{1+\beta u_i}, \quad g(u_i, v_i)=\frac{\beta u_i v_i}{1+\beta u_i}-\frac{\gamma v_i^2}{e+\eta v_i}.
$$
In our model, the prey population $u_i$ and predator population $v_i$ occupy every node in the network. The Interaction within each node, such as the predator-prey relationship between populations, is regarded as the reaction term in model~\eqref{self-diffusion}, and the diffusion transmission between nodes is called the diffusion term. Here, the total number of nodes in the network is $N$, and its topology is defined as a symmetric adjacency matrix whose elements $A_{i j}$ satisfy
\textcolor{red}{\begin{equation}
    A_{i j}= \begin{cases}1, & \text{if the nodes}~i~\text{and}~j~\text{are connected}~\text{where}~i, j=1, \ldots, N, \text{and}~i\neq j,\\
0, & \text{otherwise}.\end{cases}
\end{equation}}
Similar to~\cite{nakao2010turing}, here we define the degree of node $i$ as given by $k_i=\sum_{j=1}^N A_{i j}$. To meet the condition $k_1 \geq k_2 \geq \cdots k_N$ is satisfied, we sort network nodes $i$ in decreasing order of their degrees $k_i$. The diffusion of species ($u,~v$) at a node $i$ is given by the sum of the incoming fluxes from other connecting nodes $j$ to node $i$. According to Fick's law, the flux is proportional to the concentration difference between nodes. Then consider the network Laplacian matrix $L_{i j}=A_{i j}-k_i \delta_{i j}$, the diffusive flux of prey population $u$ to node $i$ is expressed as $\sum_{j=1}^N L_{i j} u_j$ and the diffusive flux of predator population $v$ to node $i$ is expressed as $\sum_{j=1}^N L_{i j} v_j$. \textcolor{red}{The biological significance represented by the parameters in model~\eqref{self-diffusion} can be found in Table~\ref{tab1}.}
\begin{table}[h]
    \centering
    \caption{\textcolor{red}{Biological significance of parameters in model~\eqref{self-diffusion}}}
    \label{tab1}
    \begin{tabular}{c|c}
        \hline Parameter & \multicolumn{1}{c}{ Biological significance } \\
        \hline
        $d_1$ & The diffusion rate of prey species \\
        $d_2$ & The diffusion rate of predator species\\
        $\alpha$ & The proportion of intrinsic growth rate to the birth rate of predators \\
        $A$ & The weak Allee effect constant\\
        $\beta$ & The environmental capacity to prey density at half-saturation \\
        $e$ & The water's light attenuation coefficients\\
        $\eta$ & The self-shading coefficients of light attenuation\\
        $\gamma$ & The mortality rate of predator species\\
        $L_{ij}$ & The elements in the Laplace matrix $L$ of network\\
\hline
    \end{tabular}
\end{table}

The structure of this paper is as follows. In Sect.~\ref{section2}, with the help of Turing stability theory, we analyze the conditions of Turing instability region and use two sets of examples to verify our theoretical analysis. In Sect.~\ref{section3}, numerical experiments are carried out on pattern formation in continuous medium (model~\eqref{liu_model}) and discrete medium (model~\eqref{self-diffusion}) respectively. In Sect.~\ref{section4}, we discuss and analyze the results obtained in this paper.
\section{Turing instability analysis}\label{section2}
This section mainly discusses the Turing instability of model~\eqref{self-diffusion}. With the help of the theory of reaction-diffusion model in continuous space, it is necessary to ensure that the positive equilibrium of model~\eqref{self-diffusion} is locally stable when there is no diffusion. This requires us first to study the stability of the positive equilibrium in the corresponding ordinary differential model.
\subsection{Stability analysis of non-diffusion model}\label{non-diffusion model}
In the beginning, we focus on the stability of the positive equilibrium of model~\eqref{self-diffusion}. Clearly, the positive equilibrium $E_*=\left(u_*, v_*\right)$ of the ordinary differential equation (ODE) or the partial differential equation (PDE) model~\eqref{self-diffusion} satisfies $f\left(u_*, v_*\right)=0$ and $g\left(u_*, v_*\right)=0$ :
\begin{equation}\label{ODE}
 \left\{\begin{array}{l}
\frac{\mathrm{d} u}{\mathrm{~d} t}=\alpha u(1-u)\left(\frac{u}{u+A}\right)-\frac{\alpha u v}{1+\beta u}=0, \\
\frac{\mathrm{d} v}{\mathrm{~d} t}=\frac{\beta u v}{1+\beta u}-\frac{\gamma v^2}{e+\eta v}=0 .
\end{array}\right.
\end{equation}
To simplify the discussion, similar to~\cite{zhang2014spatio,liu2019pattern}, we will focus on $\eta=\gamma$ and $e=1$. We clearly know that model~\eqref{ODE} has two boundary equilibria $E_0=(0,0)$ and $E_1=(1,0)$. Not only that, but the model may also have some positive equilibria. Therefore, we define the positive equilibria of the model~\eqref{ODE} as $E^{(i)}_*=\left(u^{(i)}_*, v^{(i)}_*\right),~i=1,2$ and $u^{(2)}_*>u^{(1)}_*,~v^{(2)}_*>v^{(1)}_*$. In addition, when $\beta>\frac{\gamma }{\gamma-1} $, $\gamma>1$, and $\frac{\gamma}{\beta} <A<\frac{\left((\beta \gamma-\gamma-\beta)^2+4 \beta \gamma^2\right)}{4 \beta^2 \gamma}$, model~\eqref{ODE} exhibits two positive equilibria $E^{(1)}_*=\left(u^{(1)}_*, v^{(1)}_*\right)$ and $E^{(2)}_*=\left(u^{(2)}_*, v^{(2)}_*\right)$. Through simple calculation, the positive equilibria are
$$
u^{(i)}_*=\frac{\beta \gamma-\gamma-\beta \pm\sqrt{(\beta \gamma-\gamma-\beta)^2+4(\gamma-\beta A) \beta \gamma}}{2 \beta \gamma},~ v^{(i)}_*=\frac{\beta}{\gamma} u^{(i)}_*,~i=1,2.
$$
It should be noted that when $\beta>\frac{\gamma }{1-\gamma} $, $0<\gamma<1$, and $A<\frac{\gamma}{\beta}$, model~\eqref{ODE} has only one positive equilibrium $E^{(2)}_*=\left(u^{(2)}_*, v^{(2)}_*\right)$.
Let $p=\beta \gamma-\gamma-\beta$, $q=(\gamma-\beta A) \beta \gamma$, and calculate the Jacobian matrix of model~\eqref{ODE} at $E^{(i)}_*$, which is given by $J_{E^{(i)}_*}=\left[\begin{array}{ll}a_{10} & a_{01} \\ b_{10} & b_{01}\end{array}\right]$, where
\[a_{10}=\frac{\alpha v^{(i)}_*}{1+\beta u^{(i)}_*}-\frac{\alpha u^{(i)}_* v^{(i)}_*}{(1+\beta u^{(i)}_*)(1-u^{(i)}_*)}+\frac{\alpha A v^{(i)}_*}{(1+\beta u^{(i)}_*)(A+u^{(i)}_*)}-\frac{\alpha v^{(i)}_*}{(1+\beta u^{(i)}_*)^2},~a_{01}=\frac{-\alpha u^{(i)}_*}{1+\beta u^{(i)}_*},\]
\[b_{10}=\frac{\beta v^{(i)}_*}{\left(1+\beta u^{(i)}_*\right)^2},~
b_{01}=\frac{-\beta u^{(i)}_*}{\left(1+\beta u^{(i)}_*\right)^2}.\]
We can easily know that the characteristic polynomial is
$$
H^{(i)}(\lambda)=\lambda^2-T^{(i)} \lambda+D^{(i)},~i=1,2.
$$
For the positive equilibrium $E^{(1)}_*=\left(u^{(1)}_*, v^{(1)}_*\right)$,
where
$$D^{(1)}=\frac{-\alpha \beta (u^{(1)}_*)^2 v^{(1)}_*}{2 \beta
 \gamma^2 (1+\beta u^{(1)}_*)^4(1-u^{(1)}_*)(u^{(1)}_*+A)}\sqrt{p^2-4q}(2A\beta \gamma + p -\sqrt{p^2-4q})<0.$$
 Obviously, the equilibrium $E^{(1)}_*$ is unstable which is a saddle.
For the positive equilibrium $E^{(2)}_*=\left(u^{(2)}_*, v^{(2)}_*\right)$,
where
$$D^{(2)}=\frac{\alpha \beta (u^{(2)}_*)^2 v^{(2)}_*}{2 \beta
 \gamma^2 (1+\beta u^{(2)}_*)^4(1-u^{(2)}_*)(u^{(2)}_*+A)}\sqrt{p^2-4q}(2A\beta \gamma + p +\sqrt{p^2-4q})>0,$$
 and suppose there is an $\alpha_H$ that makes $T^{(2)}=0$, so we have the following equation
 $$T^{(2)}=\alpha_H(\frac{ v^{(2)}_*}{1+\beta u^{(2)}_*}-\frac{u^{(2)}_* v^{(2)}_*}{(1+\beta u^{(2)}_*)(1-u^{(2)}_*)}+\frac{A v^{(2)}_*}{(1+\beta u^{(2)}_*)(A+u^{(2)}_*)}-\frac{v^{(2)}_*}{(1+\beta u^{(2)}_*)^2})-\frac{\beta u^{(2)}_*}{\left(1+\beta u^{(2)}_*\right)^2}=0.$$
 Let $M=\frac{ v^{(2)}_*}{1+\beta u^{(2)}_*}-\frac{u^{(2)}_* v^{(2)}_*}{(1+\beta u^{(2)}_*)(1-u^{(2)}_*)}+\frac{A v^{(2)}_*}{(1+\beta u^{(2)}_*)(A+u^{(2)}_*)}-\frac{v^{(2)}_*}{(1+\beta u^{(2)}_*)^2}>0$, which can be obtained
 $\alpha_H=\frac{\beta u^{(2)}_*}{M\left (1+\beta u^{(2)}_*\right)^2}$. So we have the following conclusion: if $\alpha<\alpha_H$, then $T^{(2)}<0$ and the positive equilibrium $E^{(2)}_*=\left(u^{(2)}_*, v^{(2)}_*\right)$ is stable. If $\alpha>\alpha_H$, then $T^{(2)}>0$ and the positive equilibrium $E^{(2)}_*=\left(u^{(2)}_*, v^{(2)}_*\right)$ is unstable. Particularly, when $\alpha=\alpha_H$ Hopf bifurcation occurs since $\frac{\mathrm{d}T^{(2)}}{\mathrm{d}\alpha}=M>0$. Next, we give corresponding numerical experiments to verify our theoretical analysis.
 \subsection{Example}\label{example1}
\begin{figure*}[ht]
\centering
\subfigure[]{\includegraphics[scale=0.4]{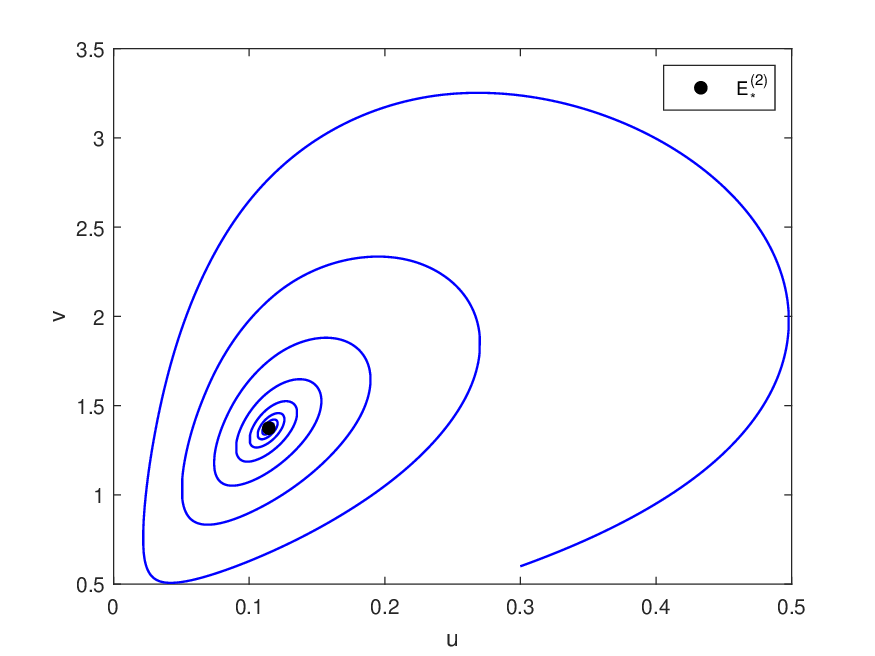}
}
\quad
\subfigure[]{\includegraphics[scale=0.4]{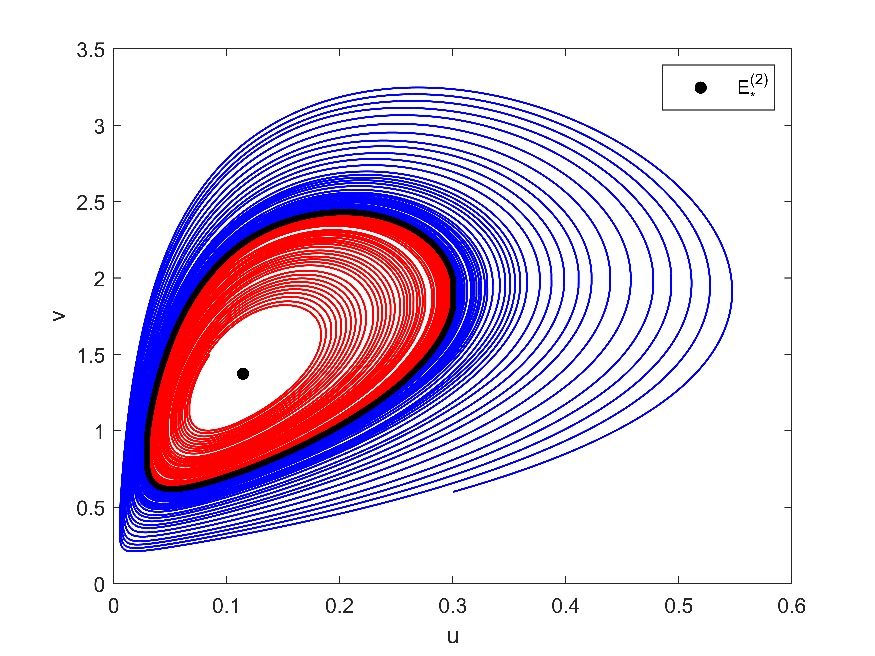}
}
\caption{(a) $E^{(2)}_*$ is stable if $0.65=\alpha<\alpha_H$ with initial values $(0.3, 0.6)$; (b) $E^{(2)}_*$ is unstable if $0.85=\alpha>\alpha_H$£¬ in particular, there is a stable limit cycle with different initial values $(0.085, 1.5)$ and $(0.3, 0.6)$.}\label{phase portraits}
\end{figure*}

 In this subsection, we provide a numerical example to illustrate the possible state of positive equilibrium $E^{(2)}_*=\left(u^{(2)}_*, v^{(2)}_*\right)$. Here, the parameters are set as $A=0.01,~\beta=6,~\gamma=0.5,~e=1,~\eta=0.5$. Therefore, model~\eqref{ODE} is in the following form:
\begin{equation}\label{Example}
 \left\{\begin{array}{l}
\frac{\mathrm{d} u}{\mathrm{~d} t}=\alpha u(1-u)\left(\frac{u}{u+0.01}\right)-\frac{\alpha u v}{1+6 u}=0, \\
\frac{\mathrm{d} v}{\mathrm{~d} t}=\frac{6 u v}{1+6 u}-\frac{0.5 v^2}{1+0.5 v}=0 .
\end{array}\right.
\end{equation}
According to the above analysis, we can know that $\alpha_H=0.827381561513881$ under this group of parameters and choose different values for $\alpha$. The phase portraits of the numerical example~\eqref{Example} are shown in Fig.~\ref{phase portraits}.
\subsection{Stability analysis of the reaction-diffusion model on network}\label{reaction-diffusion model on network}
In the case of classical continuous media, the non-uniform perturbation is usually decomposed into a set of spatial Fourier modes, representing plane waves with different wave numbers. With this idea, Othmer and Scriven noticed that the role of plane wave and wave number on the network can be reflected by the eigenvectors $\Phi^{(\alpha)}=\left(\phi^{(\alpha)}_{1}, \ldots, \phi^{(\alpha)}_{N}\right)$ and eigenvalues $\Lambda_\alpha (\alpha=1, \ldots, N)$ of their Laplace matrix, where $\sum_{j=1}^N L_{i j} \phi_j^{(\alpha)}=\Lambda_\alpha \phi_i^{(\alpha)}$~\cite{othmer1971instability}. All eigenvalues of $L_{i j}$ are non-positive real numbers. According to~\cite{nakao2010turing}, we need to meet the following condition $0=\Lambda_1 \geq \Lambda_2 \geq \cdots \geq \Lambda_N$. The eigenvectors are orthonormalized as $\sum_{i=1}^N \phi_i^{(\alpha)} \phi_i^{(\beta)}=\delta_{\alpha, \beta}$ where $\alpha, \beta=1, \ldots, N$. We introduce small perturbations $\delta u_i$ and $\delta v_i$ to the uniform state as $\left(u_i, v_i\right)=(u^{(2)}_*, v^{(2)}_*)+\left(\delta u_i, \delta v_i\right)$ and the following equation can be obtained by linearizing model~\eqref{self-diffusion}:
\begin{equation}\label{linear model}
\begin{array}{cc}
    \frac{\mathrm{d}\delta u_i}{\mathrm{d} t} =a_{10} \delta u_i+a_{01} \delta v_i+d_1 \sum_{j=1}^N L_{i j} \delta u_j,\\
    \frac{\mathrm{d}\delta v_i}{\mathrm{d} t}=b_{10} \delta u_i+b_{01} \delta v_i+d_2 \sum_{j=1}^N L_{i j} \delta v_j.
\end{array}
\end{equation}
Referring to previous work~\cite{nakao2010turing,liu2020turing}, expand the perturbations $\delta u_i$ and $\delta v_i$ in $\left(\phi^{(\alpha)}_{1}, \ldots, \phi^{(\alpha)}_{N}\right)$,
where
\begin{equation}\label{perturbations}
    \delta u_i(t)=\sum_{\alpha=1}^N c^1_\alpha  e^{\lambda_\alpha t} \phi_i^{(\alpha)},~\delta v_i(t)=\sum_{\alpha=1}^N c^2_\alpha  e^{\lambda_\alpha t} \phi_i^{(\alpha)}.
\end{equation}
Substituting Eq.~\eqref{perturbations} into Eq.~\eqref{linear model}, and using $\sum_{j=1}^N L_{i j} \phi_j^{(\alpha)}=\Lambda_\alpha \phi_i^{(\alpha)}$, we get the eigenvalue equations for each $\alpha~(\alpha=1, \ldots, N)$:
$$
\lambda_\alpha\left(\begin{array}{c}
c_\alpha^1 \\
c_\alpha^2
\end{array}\right)=\left(\begin{array}{cc}
a_{10}+d_1 \Lambda_\alpha & a_{01} \\
b_{10} & b_{01}+d_2 \Lambda_\alpha
\end{array}\right)\left(\begin{array}{c}
c^1_\alpha \\
c^2_\alpha
\end{array}\right).
$$
Further, the following characteristic polynomial can be written:
\begin{equation}\label{characteristic polynomial}
    {\lambda_\alpha}^{2}-P_1\left(\Lambda_\alpha\right) {\lambda_\alpha}+P_2\left(\Lambda_\alpha\right)=0,
\end{equation}
where
$$
\begin{aligned}
& P_1\left(\Lambda_\alpha\right)=a_{10}+b_{01}+\left(d_1+d_2\right) \Lambda_\alpha, \\
& P_2\left(\Lambda_\alpha\right)=a_{10} b_{01}-a_{01} b_{10}+a_{10} d_2 \Lambda_\alpha+b_{01} d_1 \Lambda_\alpha+d_1 d_2 \Lambda_\alpha^2 .
\end{aligned}
$$
At this time, the necessary and sufficient conditions for Turing instability can be summarized as follows according to the characteristic equation~\eqref{characteristic polynomial}:\\
(i) $a_{10}+b_{01}<0$,\\
(ii) $a_{10}b_{01}-a_{01}b_{10}>0$,\\
(iii) $a_{10} b_{01}-a_{01} b_{10}+a_{10} d_2 \Lambda_\alpha+b_{01} d_1 \Lambda_\alpha+d_1 d_2 \Lambda_\alpha^2<0$.\\
Furthermore, due to $P_1(\Lambda_\alpha)<0$ when $E^{(2)}_*=\left(u^{(2)}_*, v^{(2)}_*\right)$ is stable in model~\eqref{ODE} and $\Lambda_\alpha \leq 0$. For the reason that $P_2(\Lambda_\alpha)<0$ for some $\Lambda_\alpha$, the inequality
\begin{equation}\label{10}
a_{10}d_2+b_{01}d_1>0
\end{equation}
needs to be satisfied. Solving $P_2^{\prime}(\Lambda_\alpha)=0$ we obtain the critical Laplacian eigenvalue $\bar{\Lambda}_\alpha=-\frac{a_{10}d_2+b_{01}d_1}{2 d_1 d_2}$. Substituting this value in $P_2(\Lambda_\alpha)$ we have
\begin{equation}\label{11}
P_2(\Lambda_\alpha)_{\min }=\left(a_{10}b_{01}-a_{01}b_{10}\right)-\frac{(a_{10}d_2+b_{01}d_1)^2}{4 d_{1} d_{2}}.
\end{equation}
It follows from~(\ref{10}) and $P_2(\Lambda_\alpha)_{\min }<0$ that
\begin{equation}\label{12}
(a_{10}d_2+b_{01}d_1)>2 \sqrt{d_{1} d_{2}\left(a_{10}b_{01}-a_{01}b_{10}\right)}.
\end{equation}

\begin{remark}
It should be pointed out that there are two critical values, $a_{10}+b_{01}=0$ and $a_{10}b_{01}-a_{01}b_{10}+a_{10}d_2\Lambda_\alpha+b_{01}d_1\Lambda_\alpha+d_1 d_2 \Lambda_\alpha^2=0$. They correspond to the critical value conditions for generating Hopf bifurcation and Turing bifurcation respectively. And from $P_2(\Lambda_\alpha)=0$, we can determine $\Lambda_\alpha^{(1)}$ and $\Lambda_\alpha^{(2)}$ as
$$
\begin{aligned}
\Lambda_\alpha^{(1)} &=\frac{-(a_{10}d_2+b_{01}d_1)-\sqrt{\Delta}}{2 d_1 d_2}, \\
\Lambda_\alpha^{(2)} &=\frac{-(a_{10}d_2+b_{01}d_1)+\sqrt{\Delta}}{2 d_1 d_2}.
\end{aligned}
$$
Where $\Delta=(a_{10}d_2+b_{01}d_1)^2-4 d_1 d_2 (a_{10}b_{01}-a_{01}b_{10})$, and
Turing instability requires $\Lambda_\alpha^{(1)}<\Lambda_\alpha<\Lambda_\alpha^{(2)}$ to be maintained. Similar to the previous section, we will also give numerical examples to verify the theoretical analysis.
\end{remark}
\subsection{Example}\label{example2}
\begin{figure*}[ht]
\centering
\subfigure[]{\includegraphics[scale=0.4]{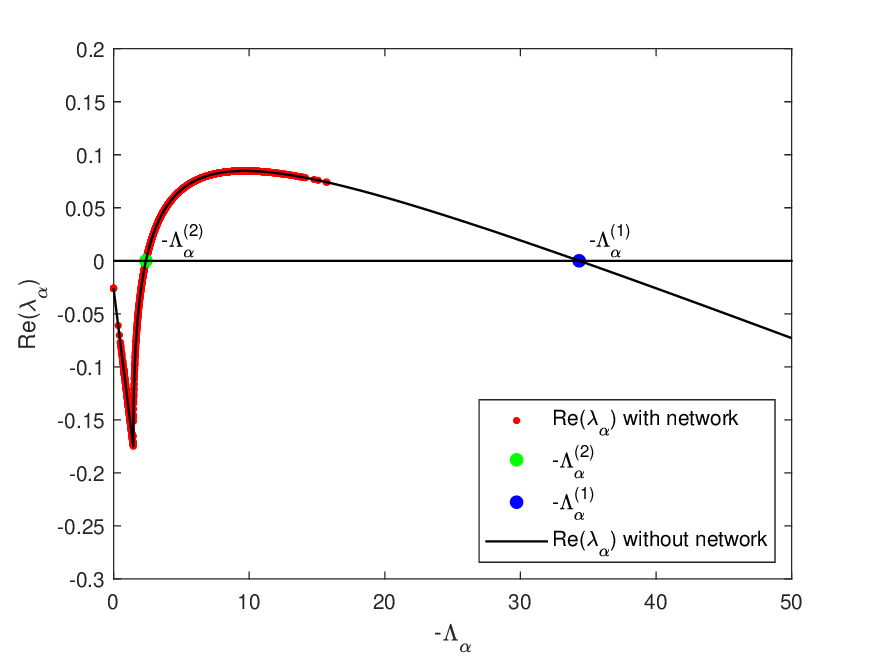}
}
\quad
\subfigure[]{\includegraphics[scale=0.4]{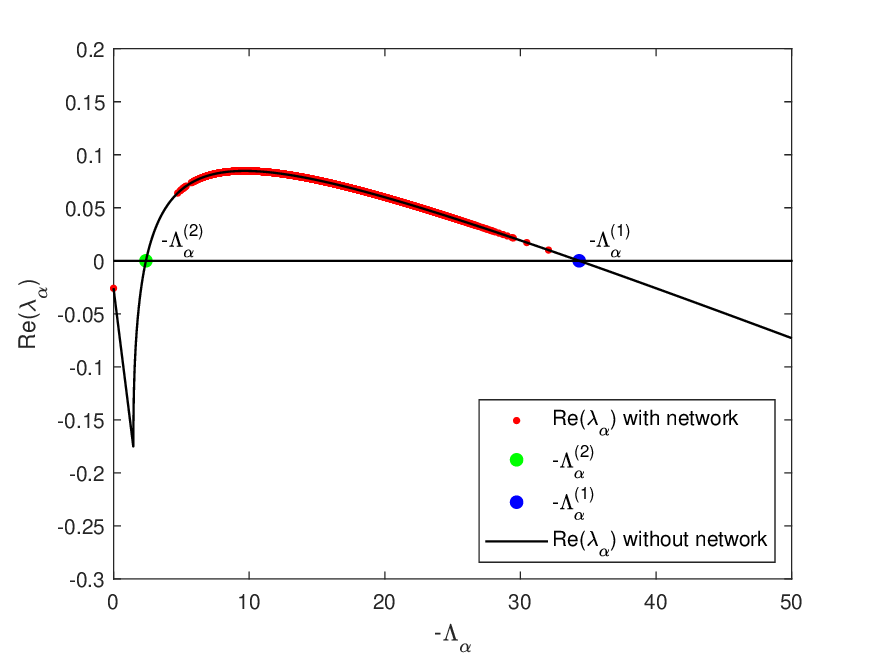}
}
\quad
\subfigure[]{\includegraphics[scale=0.4]{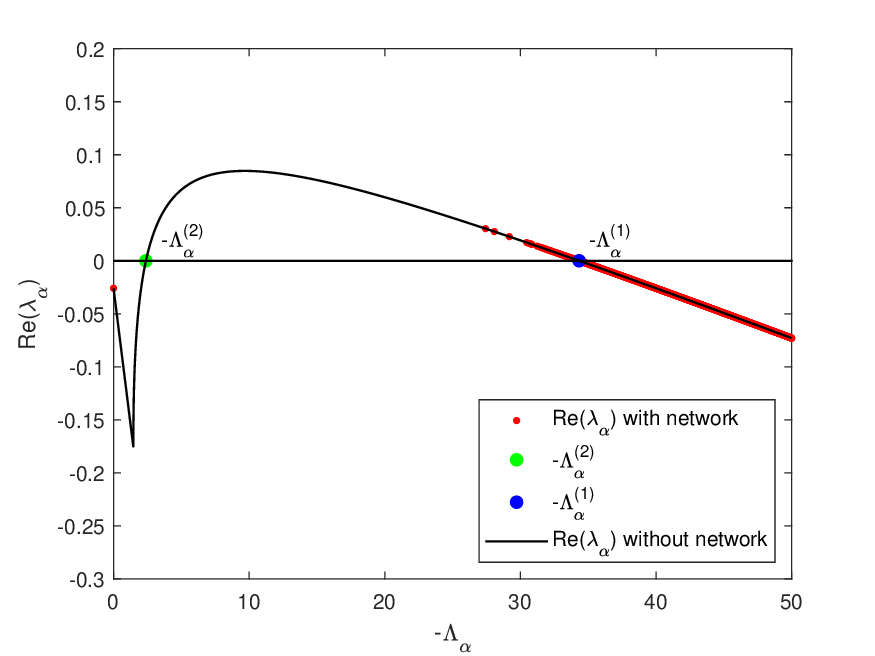}
}
\quad
\subfigure[]{\includegraphics[scale=0.4]{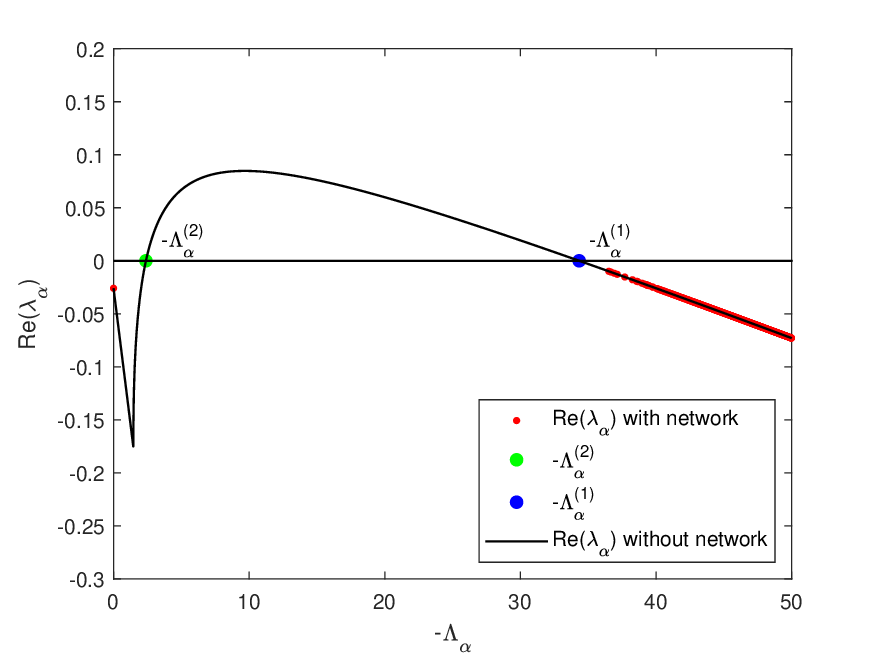}
}
\caption{Dispersion relation between Laplace eigenvalue ($\Lambda_\alpha$) and model~\eqref{self-diffusion} eigenvalue ($\lambda_\alpha$) with different average degree. (a) for average degree is $5$, (b) for average degree is $15$, (c) for average degree is $50$, and (d) for average degree is $60$.}\label{dispersion relation}
\end{figure*}

The specific theoretical analysis process has been given in subsection~\ref{reaction-diffusion model on network}, so here we only provide some numerical examples to illustrate our conclusions. Here we select parameters as $d_1=0.05$, $d_2=0.2$, and the other parameters are consistent with the parameters when $E^{(2)}_*=\left(u^{(2)}_*, v^{(2)}_*\right)$ is stable in subection~\ref{example1}. With the help of Erd\"{o}s-R\'{e}nyi network, the relationship between the real part of the eigenvalues of the model~\eqref{self-diffusion} and the eigenvalues of the Laplace matrix under different average degrees is tested. where the number of network nodes is $N=1600$. According to Eq.~\eqref{characteristic polynomial}, the relationship between Laplace eigenvalue ($\Lambda_\alpha$) and model~\eqref{self-diffusion} eigenvalue ($\lambda_\alpha$) with different average degree ($\langle k \rangle=5$, $\langle k \rangle=15$, $\langle k \rangle=50$, and $\langle k \rangle=60$) can be obtained as shown in Fig.~\ref{dispersion relation} (a), (b), (c) and (d). Obviously, when the eigenvalue of the Laplace matrix satisfies $-\Lambda_\alpha \in (-\Lambda_\alpha^{(2)}, -\Lambda_\alpha^{(1)})$, the model~\eqref{self-diffusion} placed on the average network exhibits Turing patterns. It should be noted that the Laplace eigenvalue on the network ($\Lambda_\alpha$) spectrum is discrete (red dots). In order to better understand the visualization results, we also draw the dispersion relationship (black curve) in the continuous case.
\section{Pattern formation on non-networks and networks}\label{section3}
In this section, we carry out numerical experiments on the pattern formation in non-network, i.e., with continuous media, and in the network, i.e., with discrete media. Through observation, it is found that in the model~\eqref{self-diffusion} under different parameters and different environments, the pattern types of prey population $u$ and predator population $v$ always correspond, that is, the evolution of $u$ and $v$ at each node is similar. Therefore, in the next numerical simulation, we only give the pattern formation of $v$.
\subsection{Pattern formation on non-networks}\label{Pattern formation on non-networks}
\begin{figure}[h]
\centering
\subfigure[]{\includegraphics[scale=0.3]{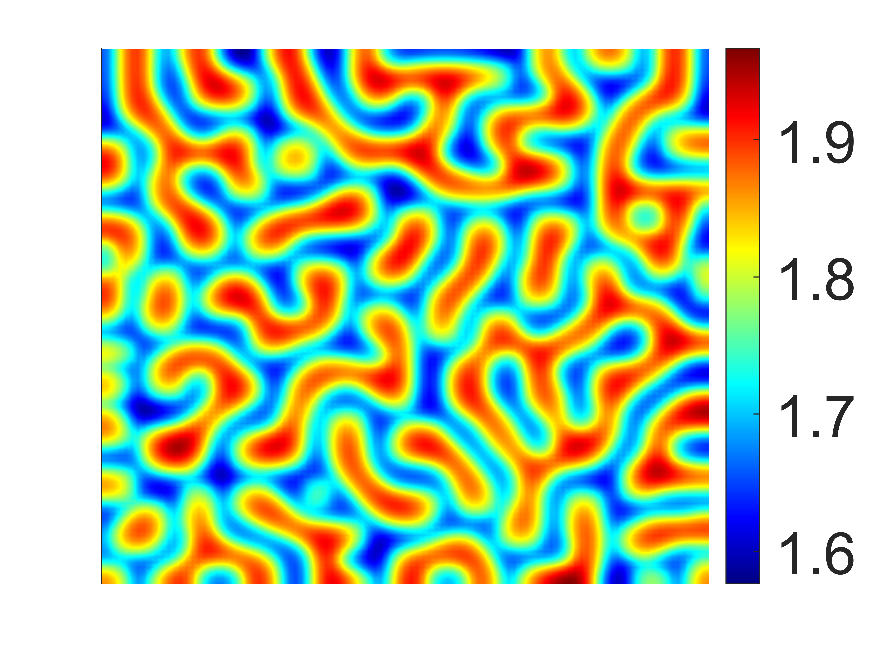}
}
\quad
\subfigure[]{\includegraphics[scale=0.3]{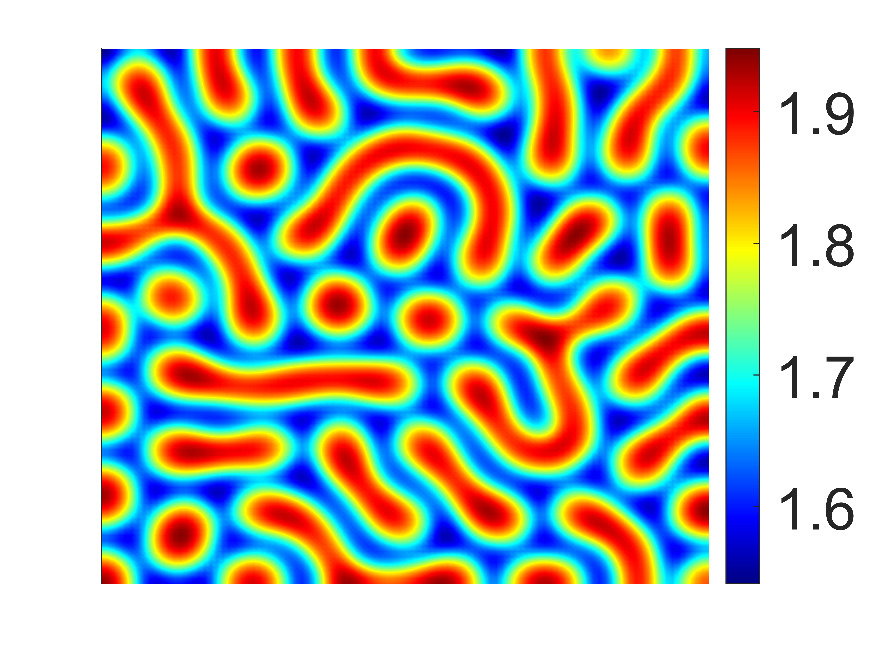}
}
\quad
\subfigure[]{\includegraphics[scale=0.3]{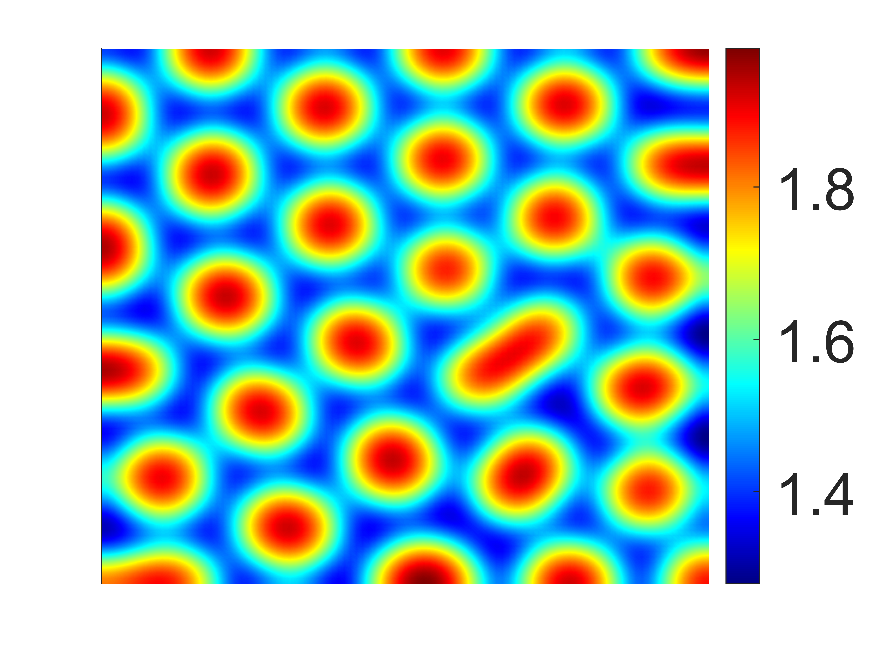}
}
\caption{Turing pattern on non-networks environment with prey diffusion coefficient is  $d_1=0.0005$ for (a),~$d_1=0.001$ for (b) and $d_1=0.005$ for (c). }\label{non-networks}
\end{figure}

\begin{figure}[h]
\centering
\subfigure[]{\includegraphics[scale=0.3]{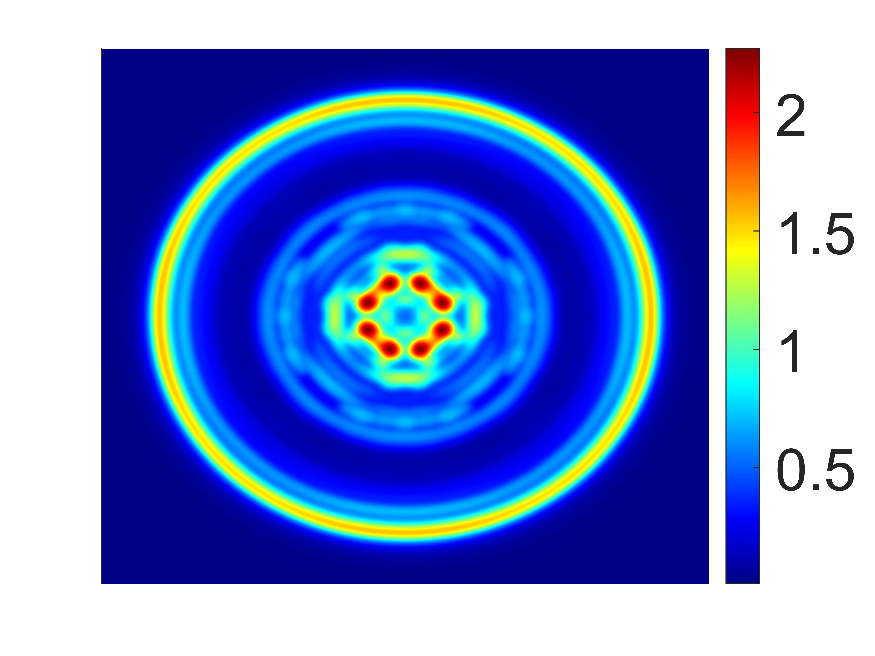}
}
\quad
\subfigure[]{\includegraphics[scale=0.3]{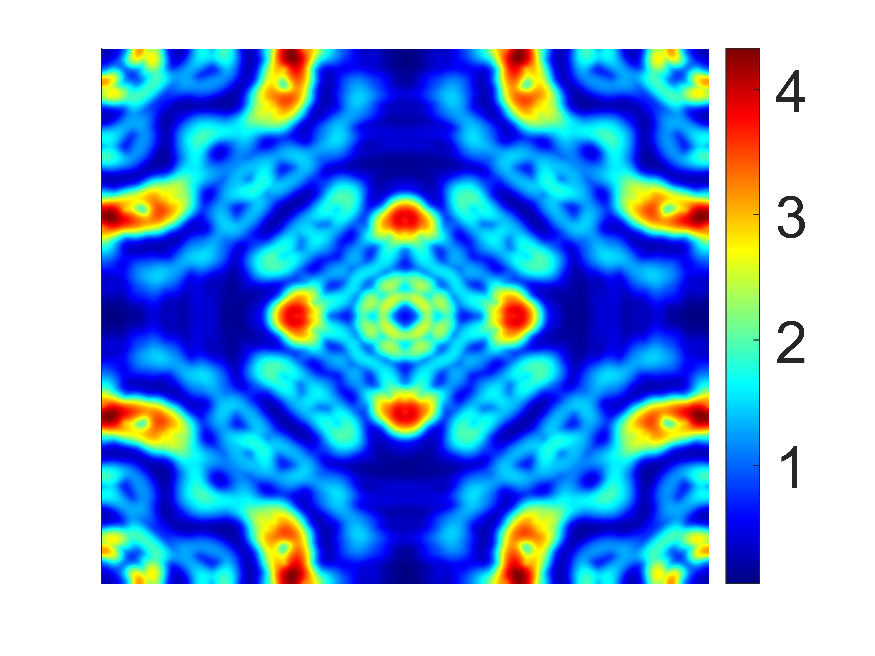}
}
\quad
\subfigure[]{\includegraphics[scale=0.3]{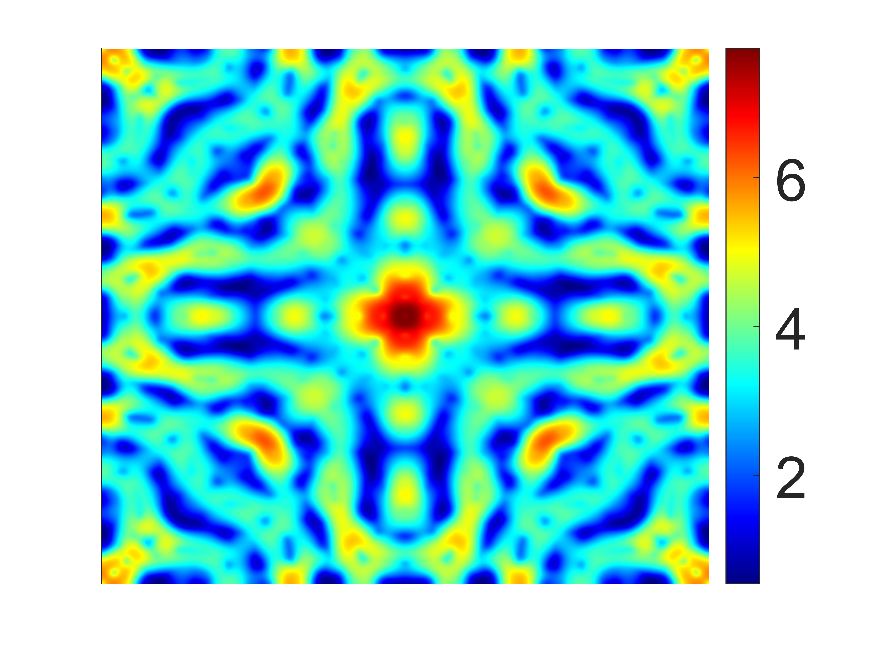}
}
\quad
\subfigure[]{\includegraphics[scale=0.3]{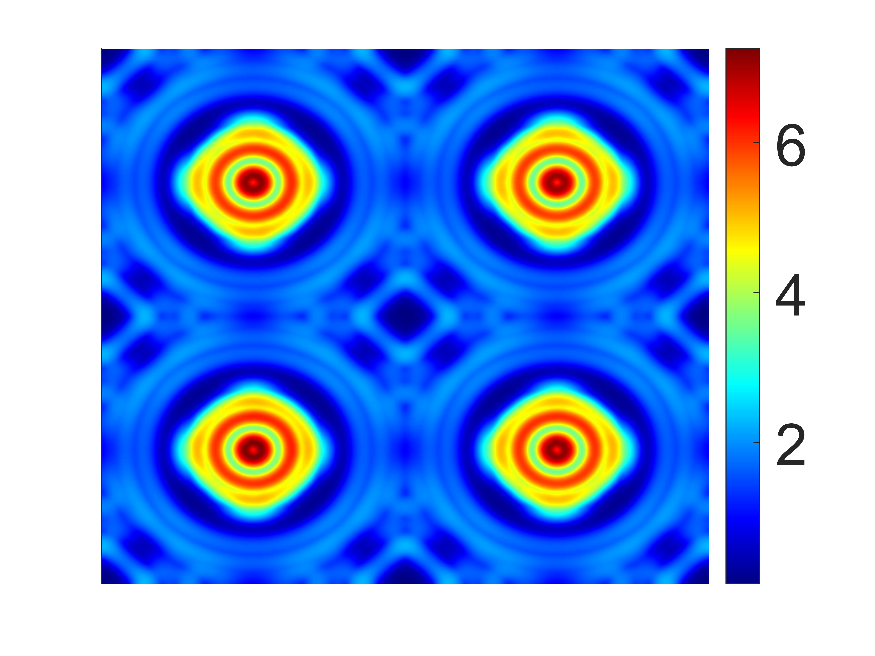}
}
\quad
\subfigure[]{\includegraphics[scale=0.3]{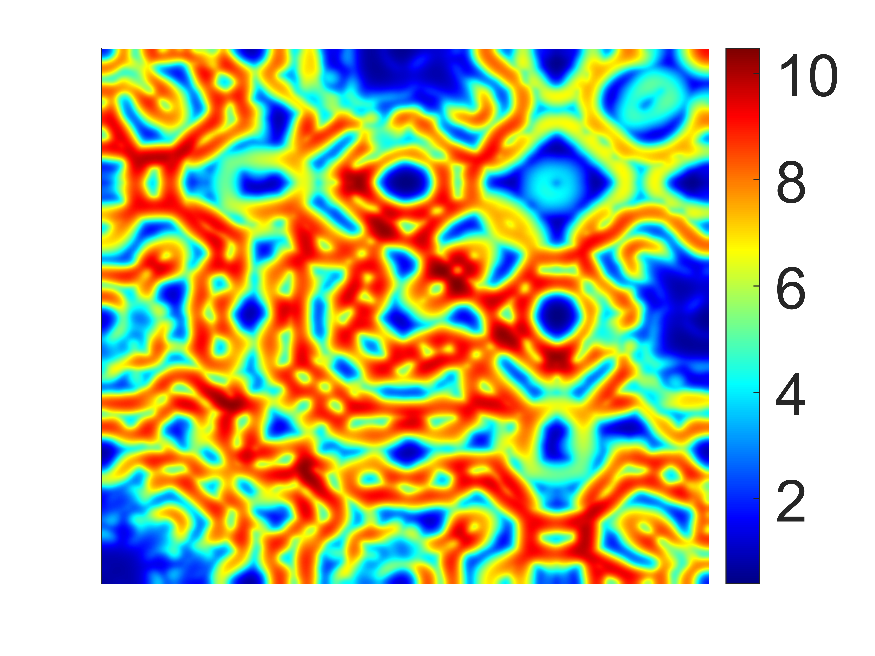}
}
\quad
\subfigure[]{\includegraphics[scale=0.3]{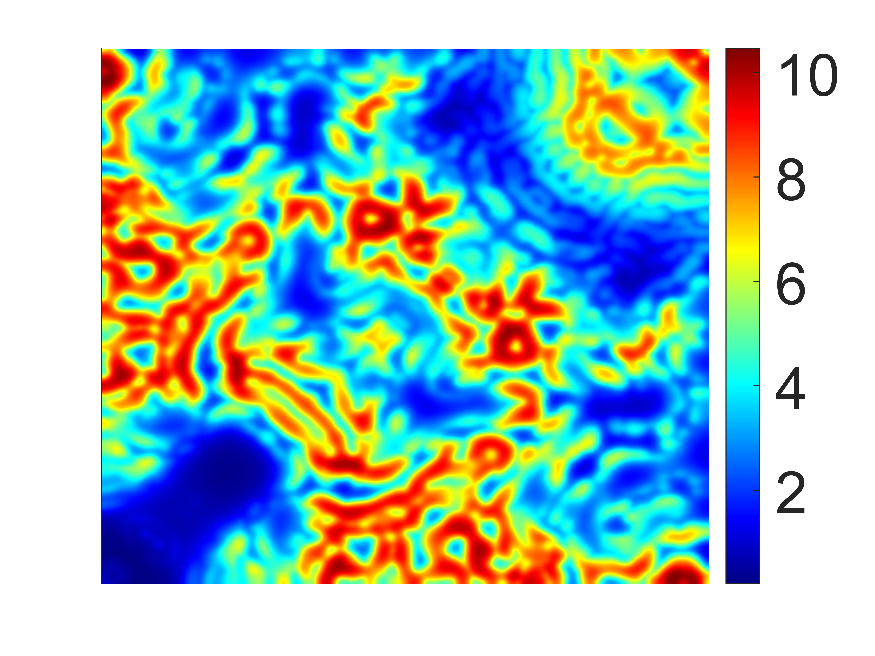}
}
\caption{Initial value induced pattern formation: From left to right, $t=500,~t=1050$, and $t=2000$. For (a), (b) and (c), $\left(u_0, v_0\right)=\left(u^{(2)}_*, v^{(2)}_*\right)-0.01 \times(1,1)$ if $(x-20)^2+(y-20)^2<50$, otherwise $\left(u_0, v_0\right)=\left(u^{(2)}_*, v^{(2)}_*\right)$. For (d), (e) and (f), $\left(u_0, v_0\right)=\left(u^{(2)}_*, v^{(2)}_*\right)-0.01 \times(1,1)$ if $(x-10)^2+(y-10)^2<20$, $(x-10)^2+(y-30)^2<20$, $(x-30)^2+(y-10)^2<20$, or $(x-30)^2+(y-30)^2<20$, otherwise $\left(u_0, v_0\right)=\left(u^{(2)}_*, v^{(2)}_*\right)$. Other parameter values are: $d_1 =0.001,~d_2 =0.1,~A=0.01,~\alpha=0.3,~\beta=14,~\gamma=0.7,~e=0.1$, and~$\eta=1.$}\label{Initial value induced pattern}
\end{figure}

In this subsection, we first show the pattern formation of model ~\eqref{self-diffusion} in continuous media under the control of parameters and initial values. Here we use the forward Euler method as our main numerical method, in which the time interval is defined as $\Delta t=0.01$, the total time is defined as $T=2000$, the space interval is defined as $h=0.1$, the 2D (two-dimensional) simulation regions are defined as $\Omega=[0,10] \times[0,10]$ in Fig.~\ref{non-networks} and $\Omega=[0,40] \times[0,40]$ in Fig.~\ref{Initial value induced pattern} under Neumann boundary conditions (zero-flux boundary), and for Fig.~\ref{non-networks}, we apply a small perturbation near the equilibrium point $E^{(2)}_*=\left(u^{(2)}_*, v^{(2)}_*\right)$ as our initial value expressed as $$
\begin{aligned}
& u(0, x, y)=\left(u^{(2)}_*\right)+0.1 \times \operatorname{rand}(0,1), \\
& v(0, x, y)=\left(v^{(2)}_*\right)+0.1 \times \operatorname{rand}(0,1),
\end{aligned}
$$
where random small perturbations are generated using ``rand'' function. Therefore, the Laplacian (i.e., diffusion term) in the standard five-point explicit finite difference scheme can be expressed as:
$$
\begin{aligned}
\Delta_{h} u_{i, j}^n & =\frac{u_{i+1, j}^n+u_{i-1, j}^n+u_{i, j+1}^n+u_{i, j-1}^n-4 u_{i, j}^n}{h^2}, \\
\Delta_{h} v_{i, j}^n & =\frac{v_{i+1, j}^n+v_{i-1, j}^n+v_{i, j+1}^n+v_{i, j-1}^n-4 v_{i, j}^n}{h^2},
\end{aligned}
$$
where $i$ and $j$ represent the position in the grid, and $n=1,\dots,N,~N=T/\Delta t$ represents the number of iterations. Therefore, we try to verify the test model without considering the network. The purpose is to explore the influence of different control parameters and initial values on the pattern formation of model~\eqref{liu_model} in continuous media. First, we select appropriate parameters to satisfy the Turing instability region in Fig.~\ref{dispersion relation}, where the predator diffusion rate is $d_2=0.2$, and the other parameters are consistent with the parameters in Fig.~\ref{phase portraits} (a). Then through extensive numerical simulation, we obtained three different types of patterns corresponding to different prey diffusion coefficients ($d_1=0.0005,~d_1=0.001$ and $d_1=0.005$): labyrinthine pattern, the mixture of hot spot and labyrinthine pattern, and hot spot pattern as shown in Fig.~\ref{non-networks}. Inspired by~\cite{nagano2012phase,zhang2014spatio}, we found that the initial value will also affect the pattern formation, so we made some attempts, that is, to observe the change in the pattern formation by selecting different initial values. It turns out that under different initial values, the system will have some interesting patterns, as shown in Fig.~\ref{Initial value induced pattern}.
\begin{figure}[ht]
\centering
\subfigure[]{\includegraphics[scale=0.45]{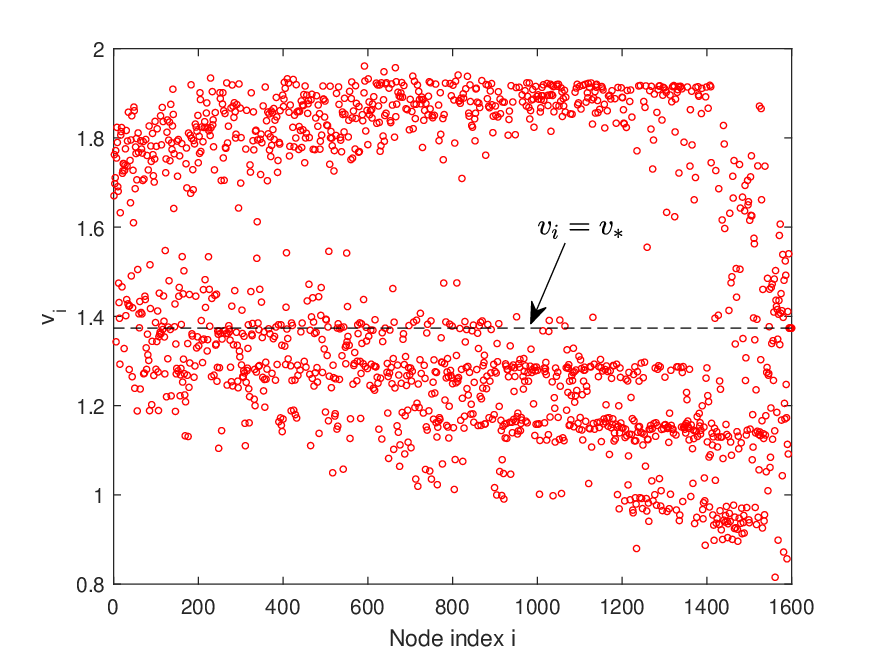}
}
\quad
\subfigure[]{\includegraphics[scale=0.45]{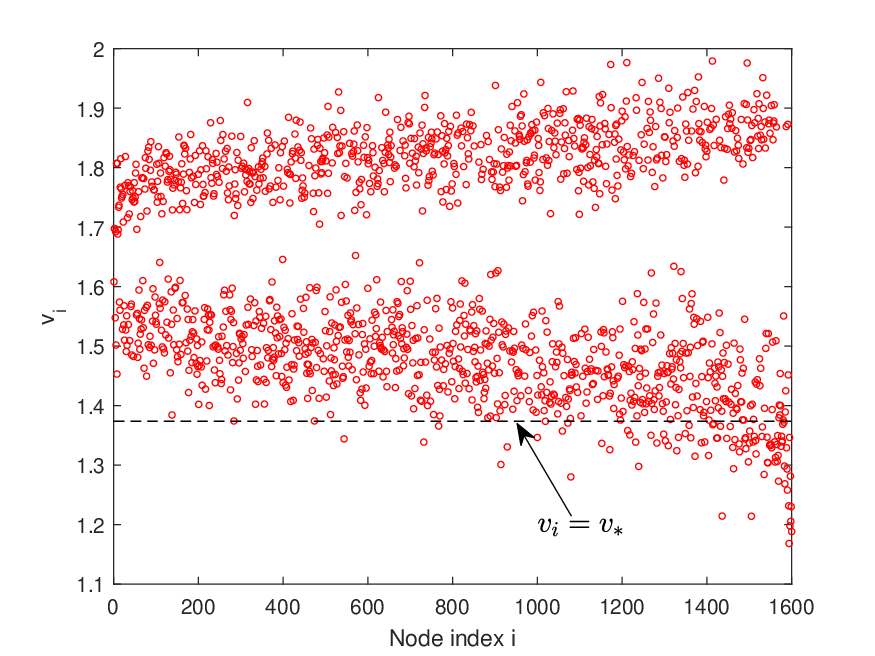}
}
\quad
\subfigure[]{\includegraphics[scale=0.45]{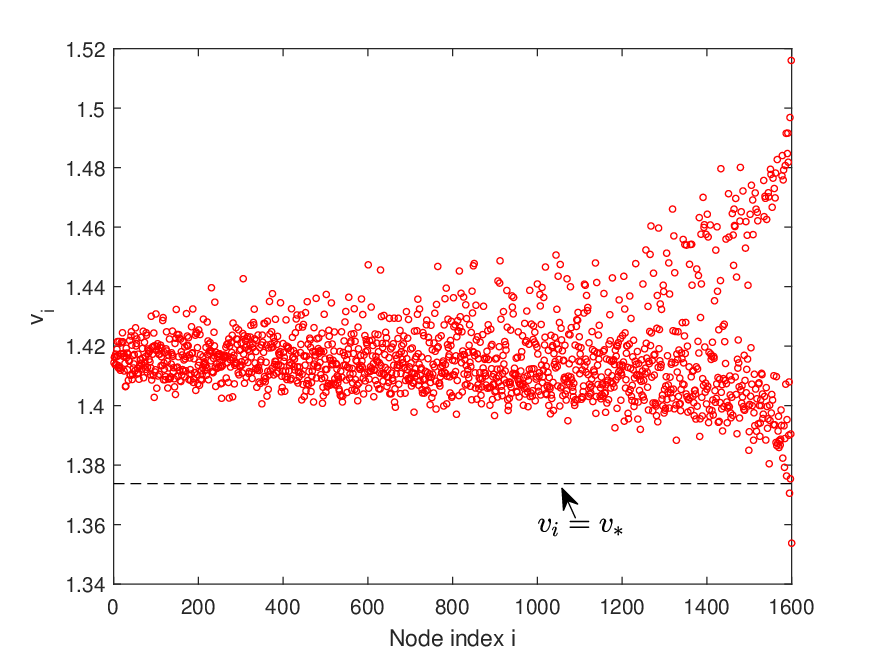}
}
\quad
\subfigure[]{\includegraphics[scale=0.45]{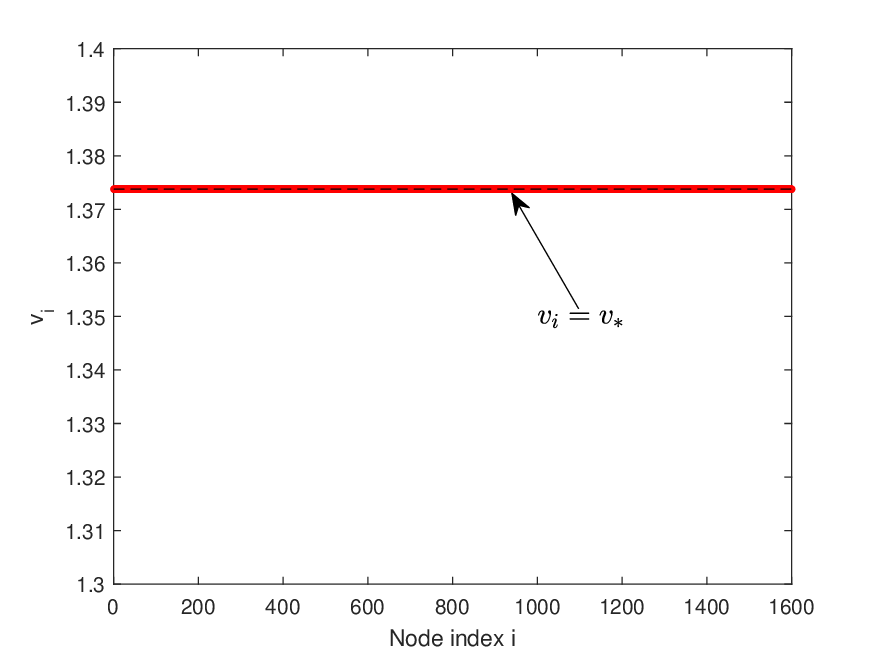}
}
\quad
\caption{\textcolor{red}{The density (red dots) of predator population ($v_i$) with four different network average degrees, where the average degree is $5$ for (a),~the average degree is $15$ for (b),~the average degree is $50$ for (c), and the average degree is $60$ for (d). The abscissa represents the ordinal number of the node index $i$ from small to large, and the ordinate represents the density value of the predator population $v_i$.}}\label{1D}
\end{figure}
\subsection{Pattern formation on networks}\label{Pattern formation on networks}
\begin{figure}[ht]
\centering
\subfigure[]{\includegraphics[scale=0.45]{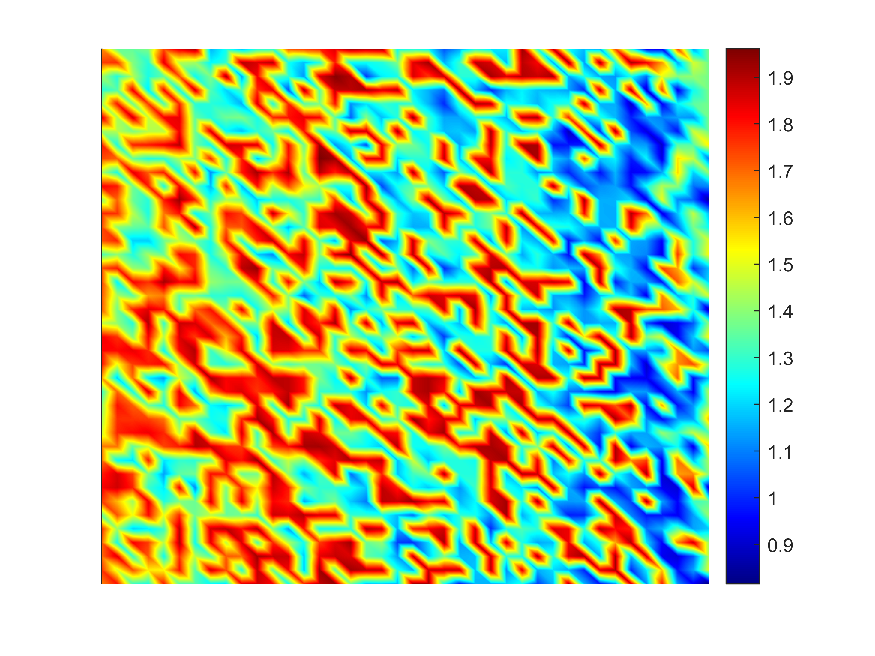}
}
\quad
\subfigure[]{\includegraphics[scale=0.45]{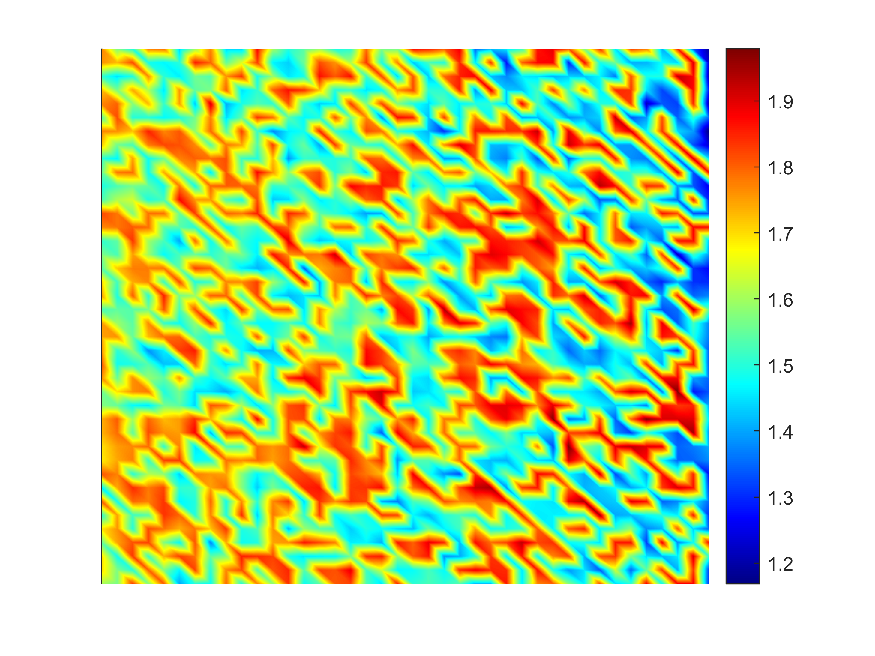}
}
\quad
\subfigure[]{\includegraphics[scale=0.45]{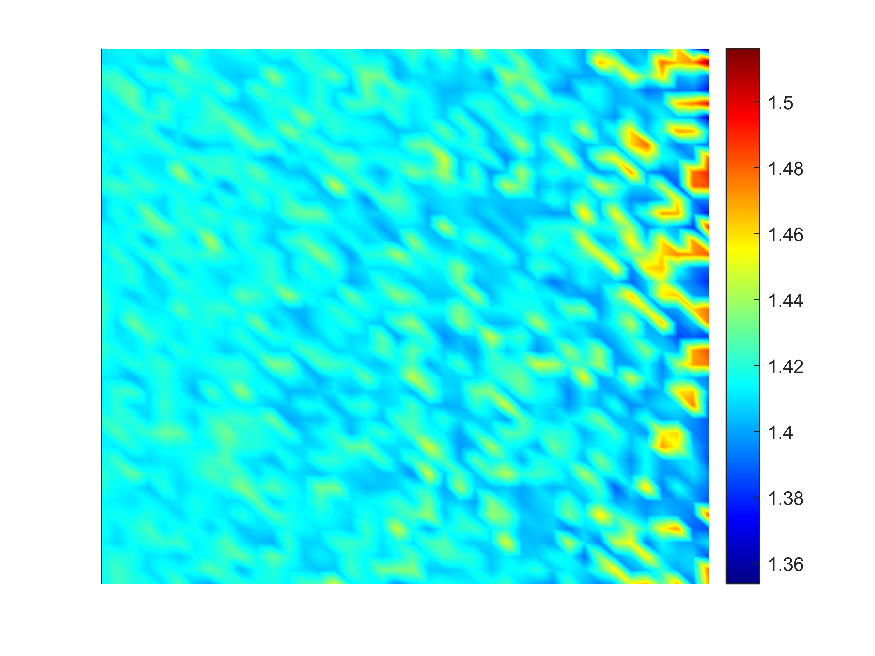}
}
\quad
\subfigure[]{\includegraphics[scale=0.45]{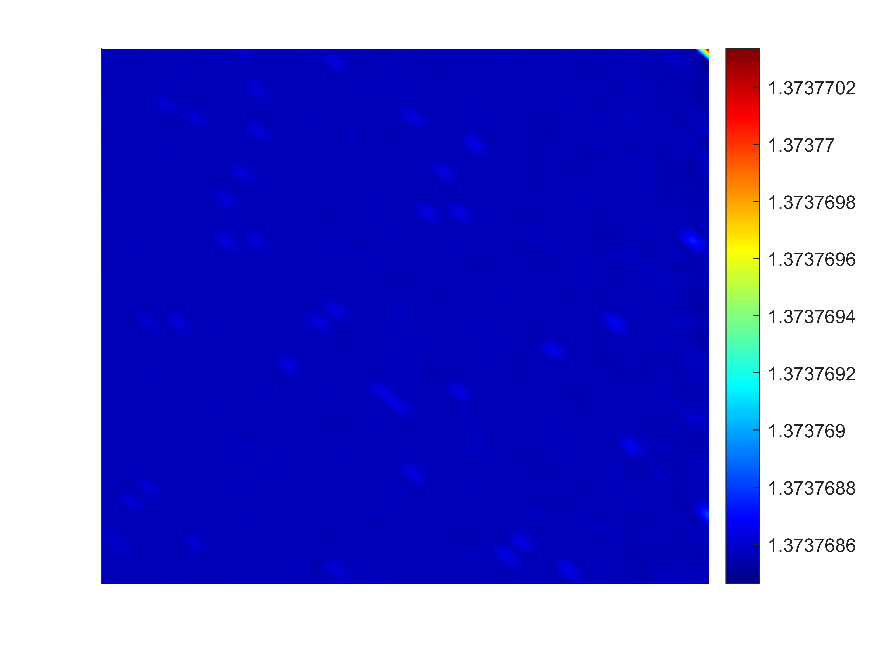}
}
\quad
\caption{Evolution of 2D (two-dimensional) \textcolor{red}{Turing patterns} on ER  random network with four different network average degrees, where the average degree is $5$ for (a),~the average degree is $15$ for (b),~the average degree is $50$ for (c), and the average degree is $60$ for (d). The specific color represents the corresponding density of the predator population ($v_i$) according to the color bar, and the density difference in space is reflected by the color difference.}\label{2D}
\end{figure}
\begin{figure}[ht]
\centering
\subfigure[]{\includegraphics[scale=0.45]{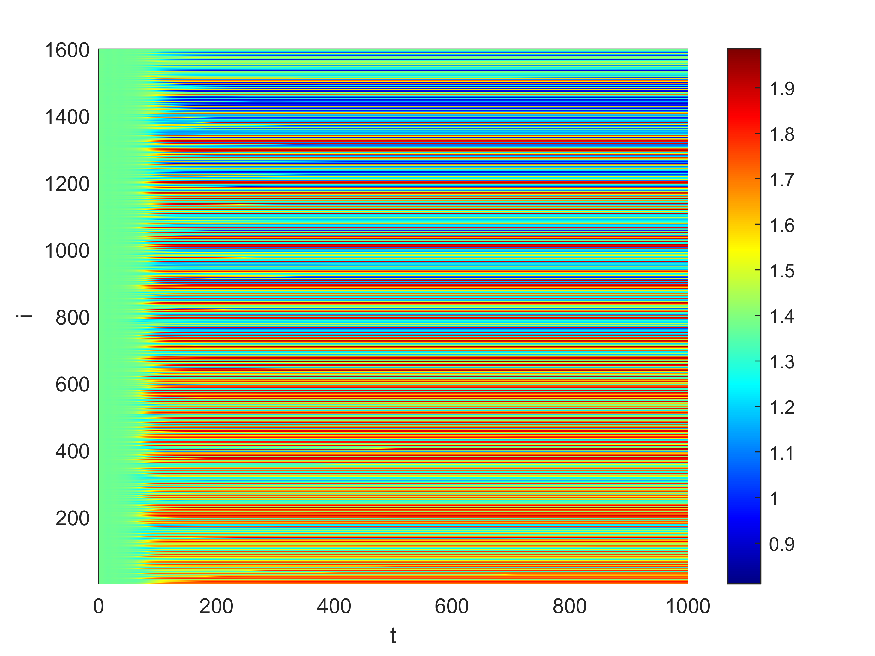}
}
\quad
\subfigure[]{\includegraphics[scale=0.45]{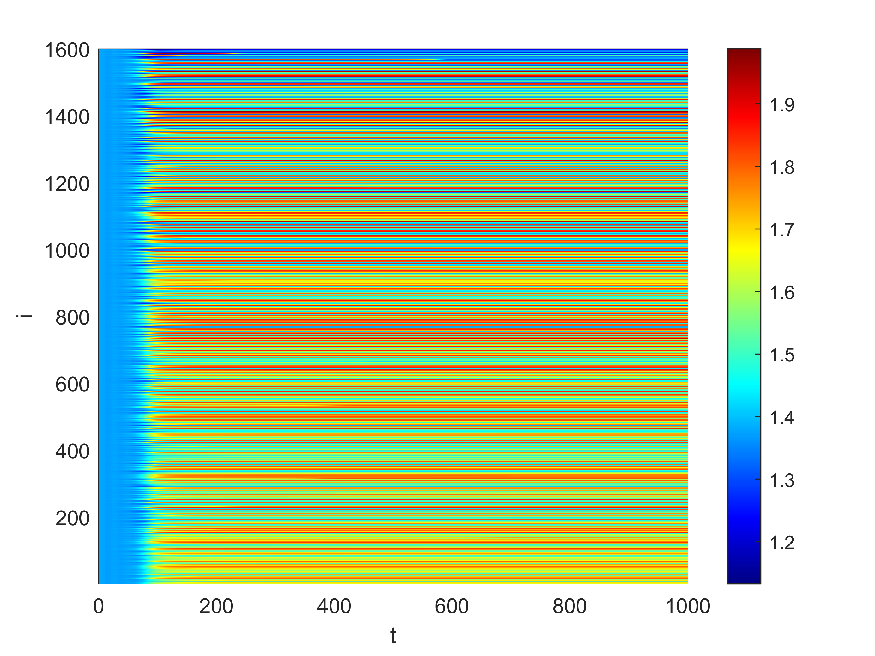}
}
\quad
\subfigure[]{\includegraphics[scale=0.45]{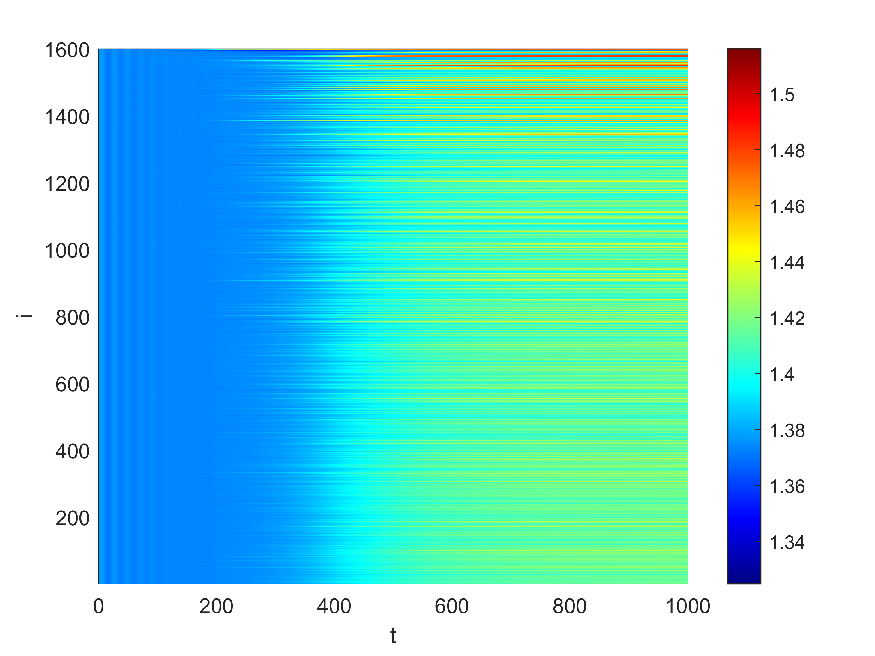}
}
\quad
\subfigure[]{\includegraphics[scale=0.45]{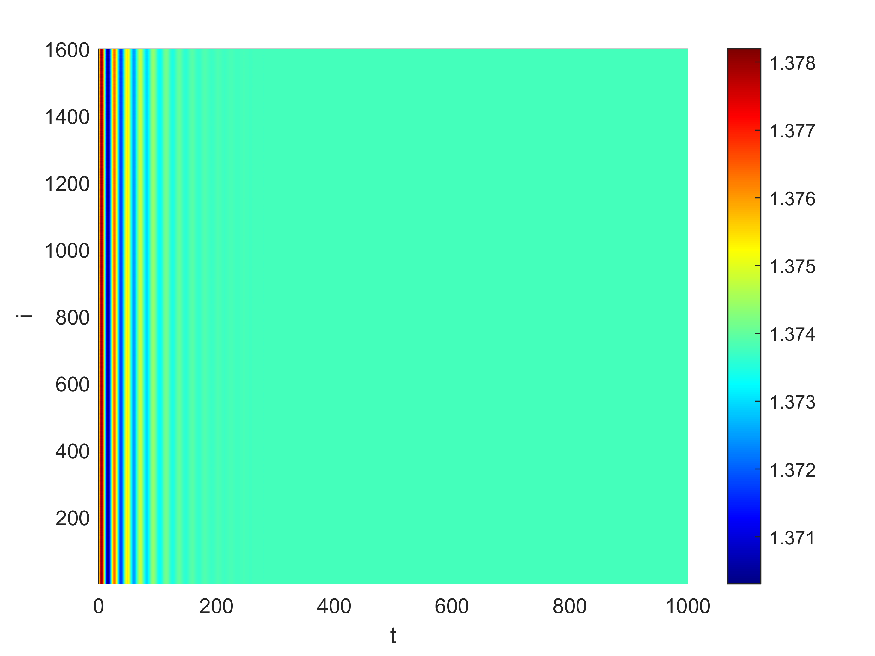}
}
\quad
\caption{The density of the predator population ($v_i$) on the ER (Erd\"{o}s-R\'{e}nyi) random network varies with time under different network average degrees, where the average degree is $5$ for (a),~the average degree is $15$ for (b),~the average degree is $50$ for (c), and the average degree is $60$ for (d). The abscissa represents the time $t$ from small to large, and the ordinate represents the ordinal number of the node index $i$. The specific color represents the corresponding density of the predator population ($v_i$) according to the color bar, and the density difference in space is reflected by the color difference.}\label{shixu}
\end{figure}
\begin{figure}[ht]
\centering
\subfigure[]{\includegraphics[scale=0.45]{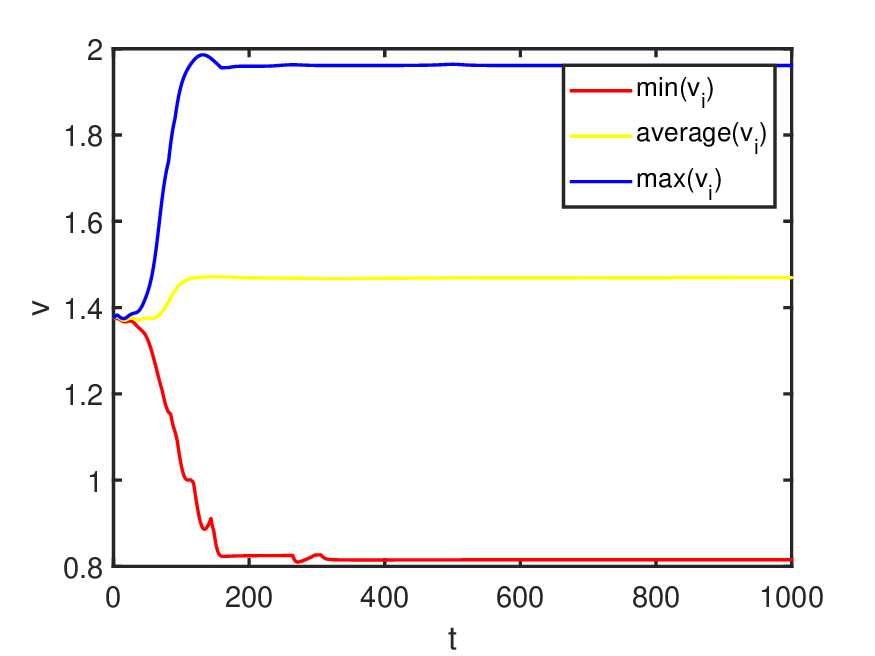}
}
\quad
\subfigure[]{\includegraphics[scale=0.45]{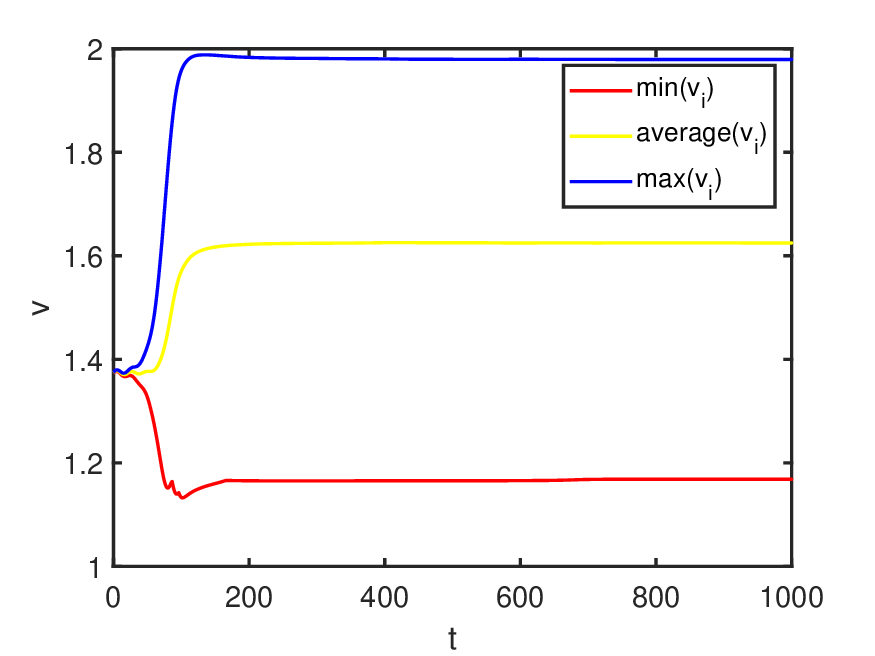}
}
\quad
\subfigure[]{\includegraphics[scale=0.45]{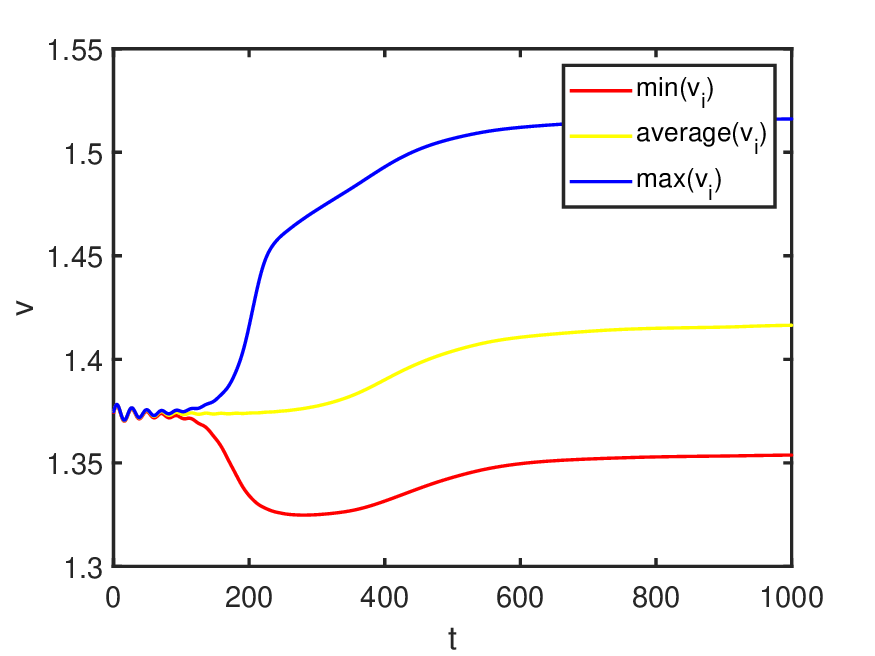}
}
\quad
\subfigure[]{\includegraphics[scale=0.45]{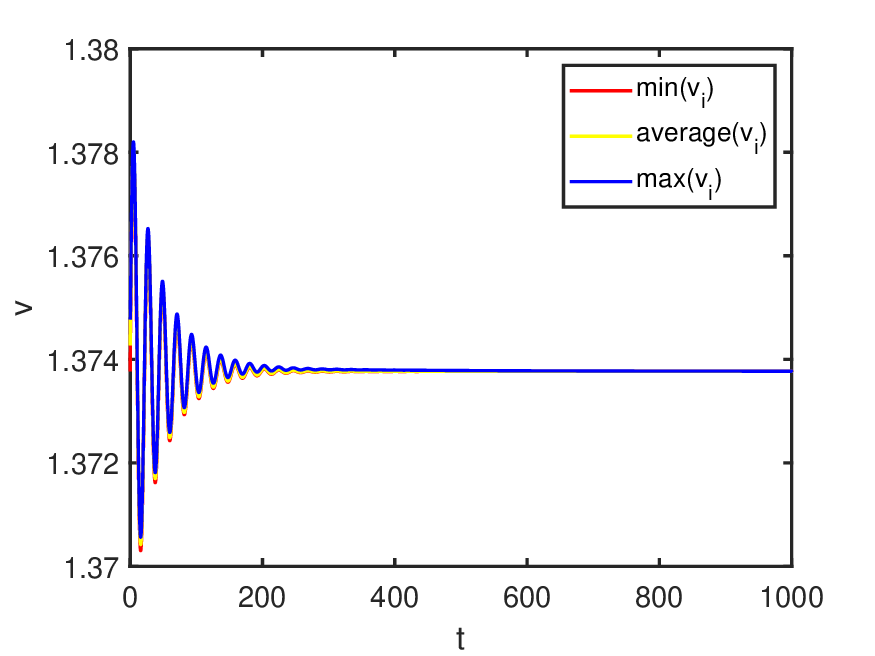}
}
\caption{The curves of the maximum, minimum, and average values of the predator population density ($v_i$) in all nodes on the network at different average degrees over time. The average degree is $5$ for (a), the average degree is $15$ for (b), the average degree is $50$ for (c), and the average degree is $60$ for (d).}\label{Average population}
\end{figure}
\begin{figure}[ht]
\centering
\subfigure[]{\includegraphics[scale=0.3]{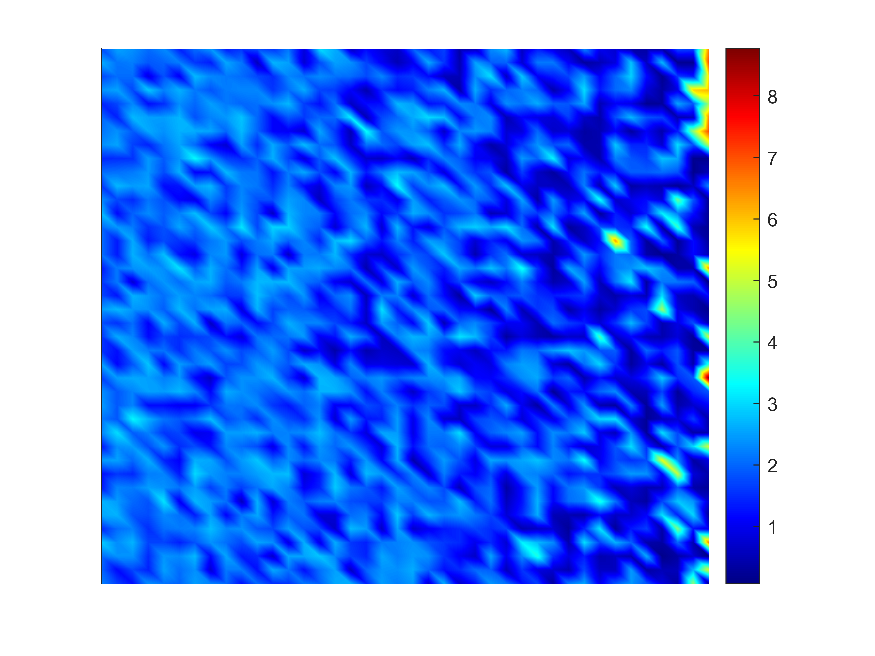}
}
\quad
\subfigure[]{\includegraphics[scale=0.3]{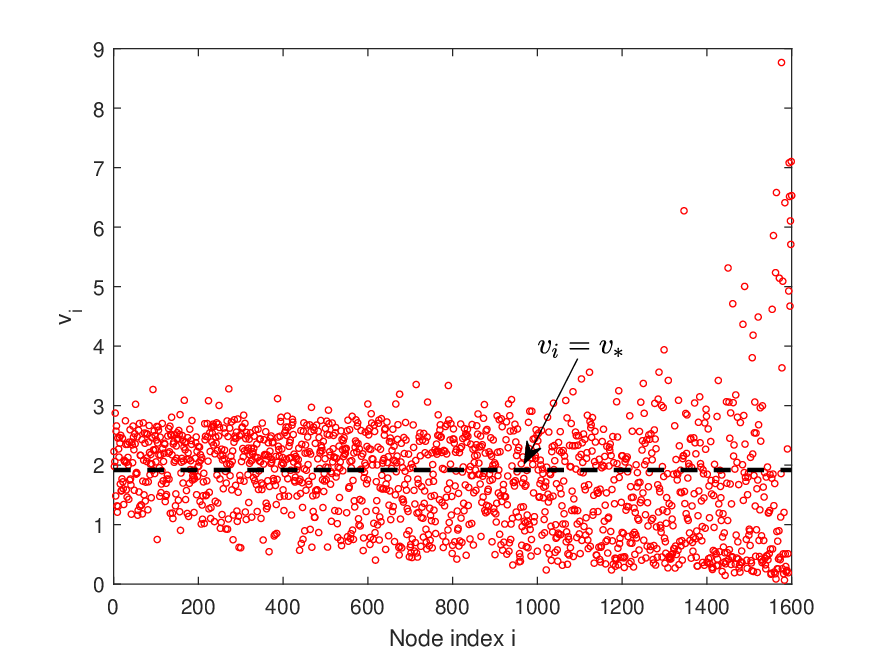}
}
\quad
\subfigure[]{\includegraphics[scale=0.3]{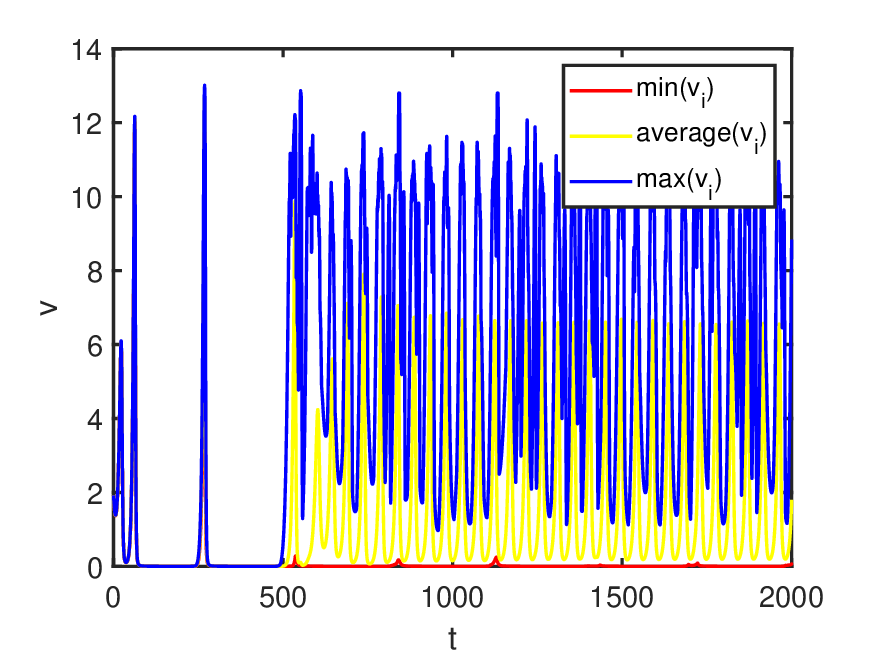}
}
\quad
\subfigure[]{\includegraphics[scale=0.3]{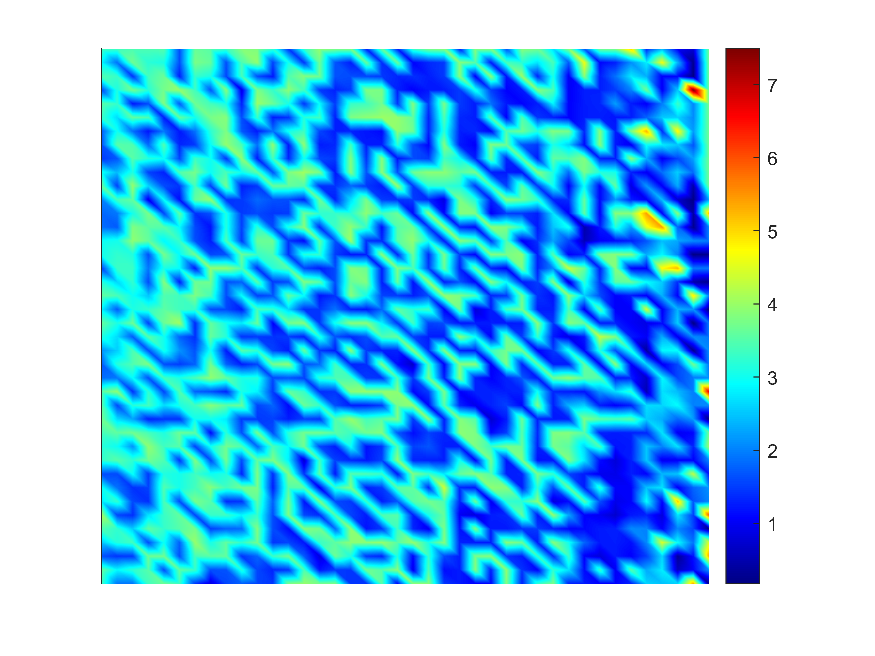}
}
\quad
\subfigure[]{\includegraphics[scale=0.3]{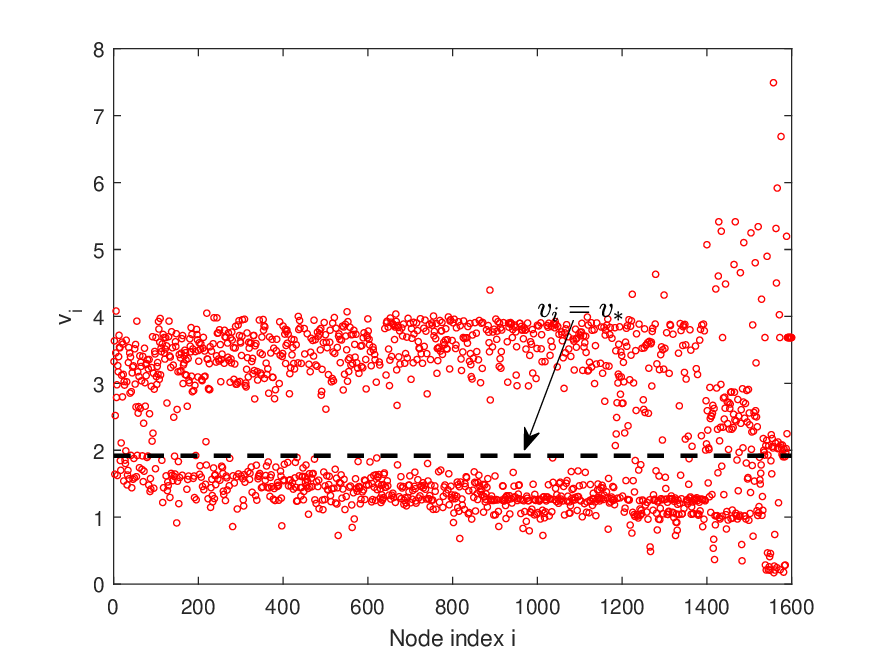}
}
\quad
\subfigure[]{\includegraphics[scale=0.3]{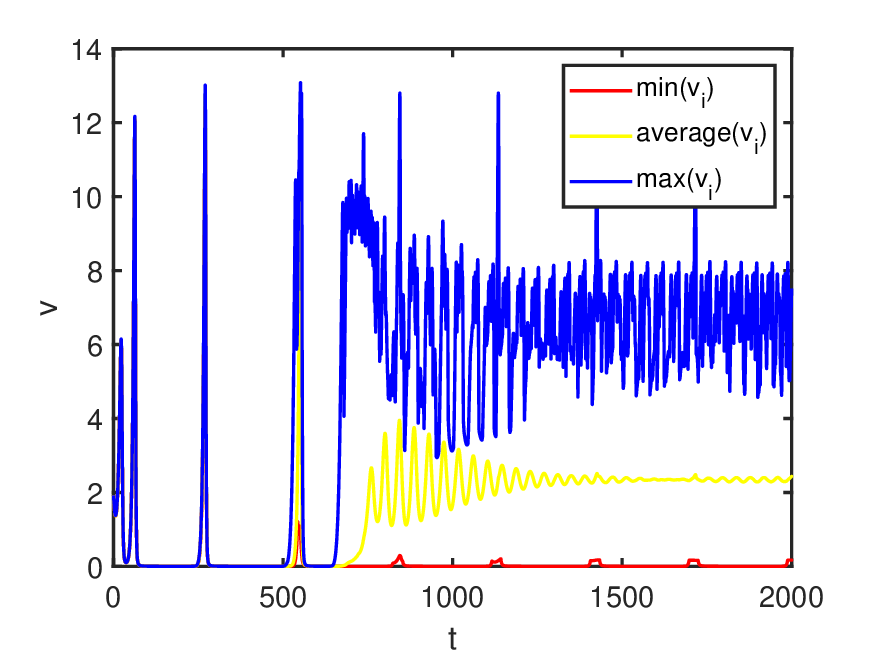}
}
\quad
\subfigure[]{\includegraphics[scale=0.3]{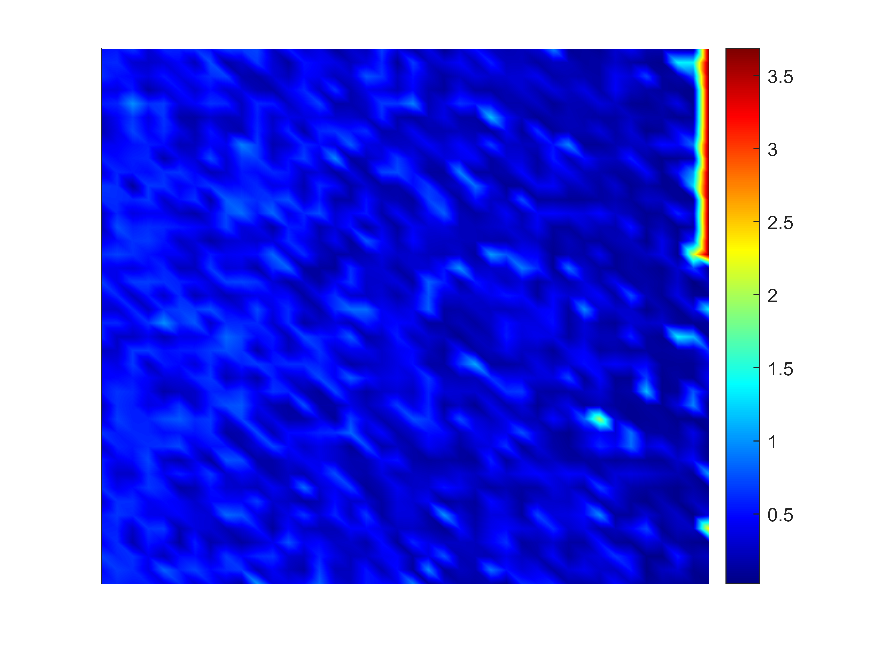}
}
\quad
\subfigure[]{\includegraphics[scale=0.3]{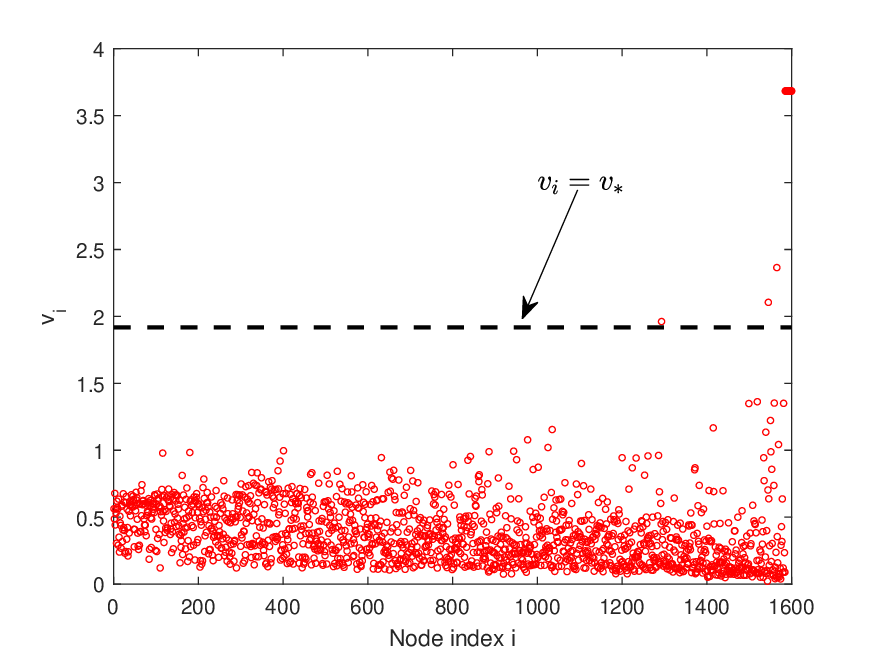}
}
\quad
\subfigure[]{\includegraphics[scale=0.3]{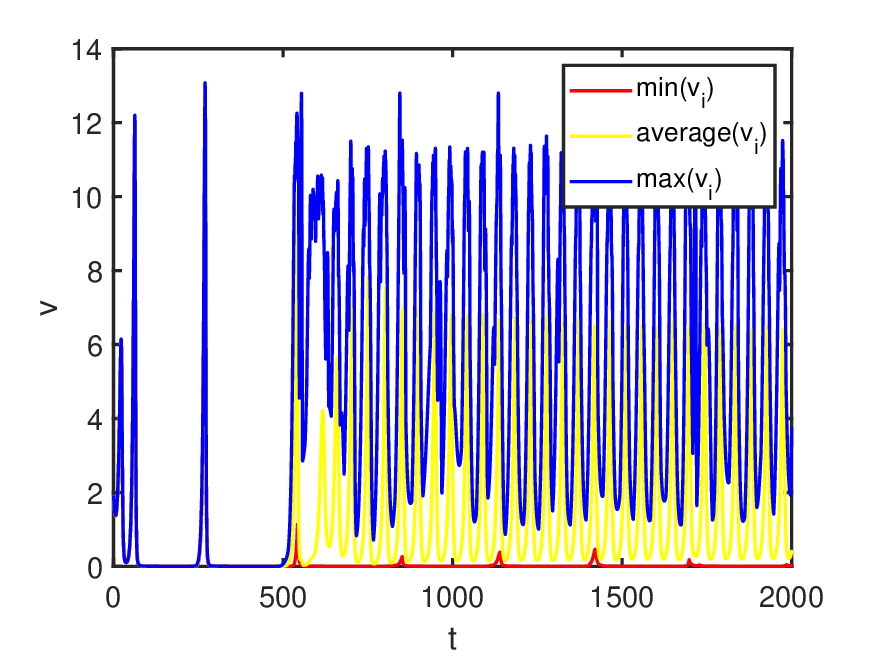}
}
\caption{\textcolor{red}{Initial value induced pattern formation on ER random network: From left to right, the evolution of 2D (two-dimensional) Turing patterns, the density of the predator population ($v_i$), and the curves of the maximum, minimum, and average values of the predator population density ($v_i$) in all nodes on ER random network. For (a), (b) and (c), $\left(u_0, v_0\right)=\left(u^{(2)}_*, v^{(2)}_*\right)+0.001 \times(rand(0,1),~rand(0,1))$. For (d), (e) and (f), $\left(u_0, v_0\right)=\left(u^{(2)}_*, v^{(2)}_*\right)+0.001 \times(1,1)$ if $(x-20)^2+(y-20)^2<10$, otherwise $\left(u_0, v_0\right)=\left(u^{(2)}_*, v^{(2)}_*\right)$. For (g), (h) and (i), $\left(u_0, v_0\right)=\left(u^{(2)}_*, v^{(2)}_*\right)+0.001 \times(1,1)$ if $(x-10)^2+(y-10)^2<10$, $(x-10)^2+(y-30)^2<10$, $(x-30)^2+(y-10)^2<10$, or $(x-30)^2+(y-30)^2<10$, otherwise $\left(u_0, v_0\right)=\left(u^{(2)}_*, v^{(2)}_*\right)$. Other parameter values are: $d_1 =0.001,~d_2 =0.1,~A=0.01,~\alpha=0.3,~\beta=14,~\gamma=0.7,~e=0.1$, ~$\eta=1.$ and $\langle k \rangle=5$.}}\label{Initial value induced pattern on network}
\end{figure}

The pattern formation in continuous media has been discussed in the previous subsection. In this subsection, we try to explore pattern formation in discrete media (complex networks). we apply a small perturbation near the equilibrium  $E^{(2)}_*=\left(u^{(2)}_*, v^{(2)}_*\right)$ as our initial value, expressed as $$
\begin{aligned}
& u_i(0)=\left(u^{(2)}_*\right)+0.001 \times \operatorname{rand}(0,1), \\
& v_i(0)=\left(v^{(2)}_*\right)+0.001 \times \operatorname{rand}(0,1),
\end{aligned}
$$
where random small perturbations are generated using ``rand'' function. In numerical experiments, it is assumed that the model is defined on the ER random network with $N = 1600$ nodes. The selection of parameters in model~\eqref{self-diffusion} is consistent with that in Fig.~\ref{non-networks} (c), in addition $E^{(2)}_*=\left(u^{(2)}_*, v^{(2)}_*\right)=(0.114480713848611, 1.373768566183330)$, $P_1(0)=-0.0517506287803922<0$, $P_2(0)=0.0820020295705703>0$, $\Lambda_\alpha^{(1)}=-34.331723013217530$, and $\Lambda_\alpha^{(2)}=-2.388520655925132$ can be calculated by substituting the set parameter values. Obviously, there exists a $P_2(\Lambda_\alpha)<0$ such that $Re(\lambda_\alpha)>0$, which satisfies the Turing instability condition as show in Fig.~\ref{dispersion relation}. Next, we design numerical experiments to study the steady-state predator density $v_i$ of each node $i$ under different network average degrees and then analyze its influence on pattern formation. Among them, Fig.~\ref{1D} shows the variation of predator population density with node index $i$ under different network average degrees. Through the observation of Fig.~\ref{1D} (a), we find that the distribution of predator populations will be divided into two groups. We define a group with high abundance as $\hat{v}_i$ and a group with low abundance as $\check{v}_i$. When the average degree of the network is $5$, we find that the distribution of predators satisfies $\check{v}_i < v^{(2)}_* < \hat{v}_i$ at this time. Still, as the average degree increases, we find some differences that the distribution of predators is more concentrated as shown in Fig.~\ref{1D} (b) and (c). Interestingly, when we increase the average degree to a certain value, such as $60$, we find that the distribution of predator populations does not differentiate at this time, and remains in a steady state as shown in Fig.~\ref{1D} (d). Correspondingly, we also give the evolution of 2D (two-dimensional) Turing patterns on ER  random network, the density of the predator population ($v_i$) on the ER  random network varies with time and the curves of the maximum, minimum, and average values of the predator population density ($v_i$) in all nodes on the network varies with time under four different network averages as shown in Fig.~\ref{2D}, Fig.~\ref{shixu}, and Fig.~\ref{Average population} (where $\langle k \rangle=5$ for (a), $\langle k \rangle=15$ for (b), $\langle k \rangle=50$ for (c), $\langle k \rangle=60$ for (d)). These figures can all illustrate the changes in the distribution patterns of the above-mentioned predator populations. Finally, we also verified the possibility of spatiotemporal patterns in ER random networks with different initial values. It was found that similar to the situation in a continuous medium, under fixed parameters, different initial values can cause differences in the spatial distribution of species, as shown in Fig.~\ref{Initial value induced pattern on network}.
\section{Conclusion}\label{section4}
We study the spatial pattern on a predator-prey model with weak Allee effect and hyperbolic mortality in continuous and discrete media. In theory, the Turing instability region in discrete media is analyzed with the help of Turing stability theory in continuous media. It can be found that the Turing instability conditions of the two are the same. The difference is that the perturbation extends to the set of Laplace matrix eigenvectors, and the dispersion relationship between the wave number and the model~\eqref{liu_model} eigenvalue $\lambda_\alpha$ in the continuous media corresponds to the dispersion relationship between the Laplace matrix eigenvalue $\Lambda_\alpha$ of the ER random network and the model~\eqref{self-diffusion} eigenvalue $\lambda_\alpha$. In continuous media, we find that the parameters and diffusion coefficient in the model control the mode generation, in which the diffusion coefficient plays a decisive role. In addition, the selection of patterns is mentioned in~\cite{zhang2014spatio,liu2019pattern}. Interestingly, when we change the initial value, many beautiful patterns appear corresponding to different initial values. Next, we try to extend consider the pattern formation on large random networks. The results show that the distribution pattern of the predator population is divided into two groups, one group is high abundance and the other group is low abundance. The stable coexistence equilibrium $E^{(2)}_*=\left(u^{(2)}_*, v^{(2)}_*\right)$ exists between the two groups. However, with the increase of the average degree of the selected network, the two groups gradually merge and then are consistent with the stable coexistence equilibrium point. At this time, there is no Turing pattern on the network, although the parameter conditions exist Turing instability region in the continuous media. Therefore, we draw the following conclusion: the average degree of the network plays an important role in the generation of the pattern, and an excessive average degree will inhibit the emergence of the Turing pattern. \textcolor{red}{Specifically, we observe that spatiotemporal patterns on discrete media (complex networks) are still influenced by initial values, and to our knowledge, this phenomenon has not been mentioned in previous work.}
\section*{CRediT authorship contribution statement}
{\bf Yong Ye:} Writing - original draft, Formal analysis, Investigation, Methodology, Software. {\bf Jiaying Zhou:} Writing - Reviewing and Editing, Supervision.
\section*{Declaration of competing interest}
All the authors declare that there is no conflict of interest during this study.
\section*{Acknowledgements}
Yong Ye acknowledges support by the scholarship from the China Scholarship Council (No. 202206120230). Jiaying Zhou acknowledges support by the scholarship from the China Scholarship Council (No. 202106120290).

\bibliography{mybibfile}

\end{document}